\documentclass[aps,twocolumn,superscriptaddress,footinbib]{revtex4}  
\usepackage{epstopdf}
\usepackage{graphicx}  
\usepackage{bm}        
\usepackage{braket}
\usepackage{exscale}                  
\usepackage[intlimits]{amsmath}       
\usepackage{amsfonts}
\usepackage{amssymb,amscd}
\usepackage{color}
\usepackage{hyperref}
\usepackage{subfigure}
\usepackage[normalem]{ulem}
\usepackage{slashed}
\usepackage{leftidx}
\usepackage{xcolor,soul}
\usepackage{txfonts}
\usepackage{charter}
\usepackage{dsfont}
\usepackage{mathrsfs}
\newtheorem{theorem}{Theorem}
\newtheorem{lemma}{Lemma}
\newtheorem{corollary}{Corollary}

\usepackage{times}
\newcommand{\op}[1]{|#1\rangle\!\langle#1|}
\newcommand{\norm}[1]{\Vert #1\Vert}

\hypersetup{colorlinks=true, citecolor=blue, urlcolor=blue, linkcolor=blue}
\begin{document}
\title{Quantum-imaginarity-based quantum speed limit}

\author{Dong-Ping Xuan}
\affiliation{School of Mathematical Sciences, Capital Normal University, Beijing 100048, China}
\author{Zhong-Xi Shen}
\affiliation{School of Mathematical Sciences, Capital Normal University, Beijing 100048, China}
\author{Wen Zhou}
\affiliation{School of Mathematical Sciences, Capital Normal University, Beijing 100048, China}
\author{Shao-Ming Fei\footnote{E-mail: feishm@cnu.edu.cn}}
\affiliation{School of Mathematical Sciences, Capital Normal University, Beijing 100048, China}
\affiliation{Max-Planck-Institute for Mathematics in the Sciences, 04103 Leipzig, Germany}
\author{Zhi-Xi Wang\footnote{E-mail: wangzhx@cnu.edu.cn}}
\affiliation{School of Mathematical Sciences, Capital Normal University, Beijing 100048, China}

\begin{abstract}
\textbf{
The quantum speed limit sets a fundamental restriction on the evolution time of quantum systems. We explore the relationship between quantum imaginarity and the quantum speed limit by utilizing measures such as relative entropy, trace distance, and geometric imaginarity. These speed limits define the fundamental constraints on the minimum time necessary for quantum systems to evolve under various dynamical processes. As applications the dephasing dynamics and dissipative dynamics are analyzed in detail. The quantum speed limit in stochastic-approximate transformations is also investigated. Our quantum speed limits provide lower bounds on how fast a physical system evolves to attain or lose certain imaginarity, with potential applications in efficient quantum computation designs, quantum control and quantum sensing.}
\end{abstract}
\maketitle

\vspace{.4cm}

\section{Introduction}\label{sect1}


The quantum speed limit (QSL) sets a fundamental constraint dictated by quantum mechanics on how fast a quantum system can evolve under a given dynamical process \cite{MT,ML,PhysRevLett.65.1697,PhysRevLett.120.070401}. It establishes a lower bound on the minimum time required for a quantum system to transition from an initial state to a final state within a specified evolution. Establishing QSLs plays a vital role in advancing quantum technologies, encompassing areas such as quantum computing \cite{NewJPhys.24.065003}, quantum metrology \cite{QuantumSciTechnol3025002}, optimal control theory \cite{PhysRevLett118100601}, and quantum thermodynamics \cite{PhysRevE97062116}. These bounds are essential for optimizing performance and efficiency in various quantum applications.


The initial  QSL was introduced for orthogonal states through the Mandelstam-Tamm (MT) bound \cite{MT}, expressed as $T_{\perp} \geq \frac{\pi}{2 \Delta E_\psi}$. Here, the energy variance $\left(\Delta E_\psi\right)^2$ is defined as $\left\langle H^2\right\rangle_\psi-\langle H\rangle_\psi^2$, with $H$ representing the Hamiltonian.
Subsequently, Margolus and Levitin (ML) established an alternative bound \cite{ML}, given by $T_{\perp} \geq \frac{\pi}{2 E_\psi}$, where the mean energy is defined as $E_\psi=\langle H\rangle_\psi-E_0$, with $E_0$ representing the ground state energy. It is worth noting that the distinction between the MT and ML bounds arises solely from the different choices of energy scaling. Generalized quantum speed limits for transitions between mixed states $\rho \rightarrow \sigma$ have been introduced in \cite{PhysRevX6021031}:
$T(\rho \rightarrow \sigma) \geq \frac{\arccos \sqrt{F(\rho, \sigma)}}{\min \left\{\Delta E_\rho, E_\rho\right\}}$, where $\sqrt{F(\rho, \sigma)}=\operatorname{tr} \sqrt{\sqrt{\rho} \sigma \sqrt{\rho}}$ is the root of fidelity between
 state $\rho$ and $\sigma$.

\begin{figure}
\centering
\includegraphics[width=1\linewidth]{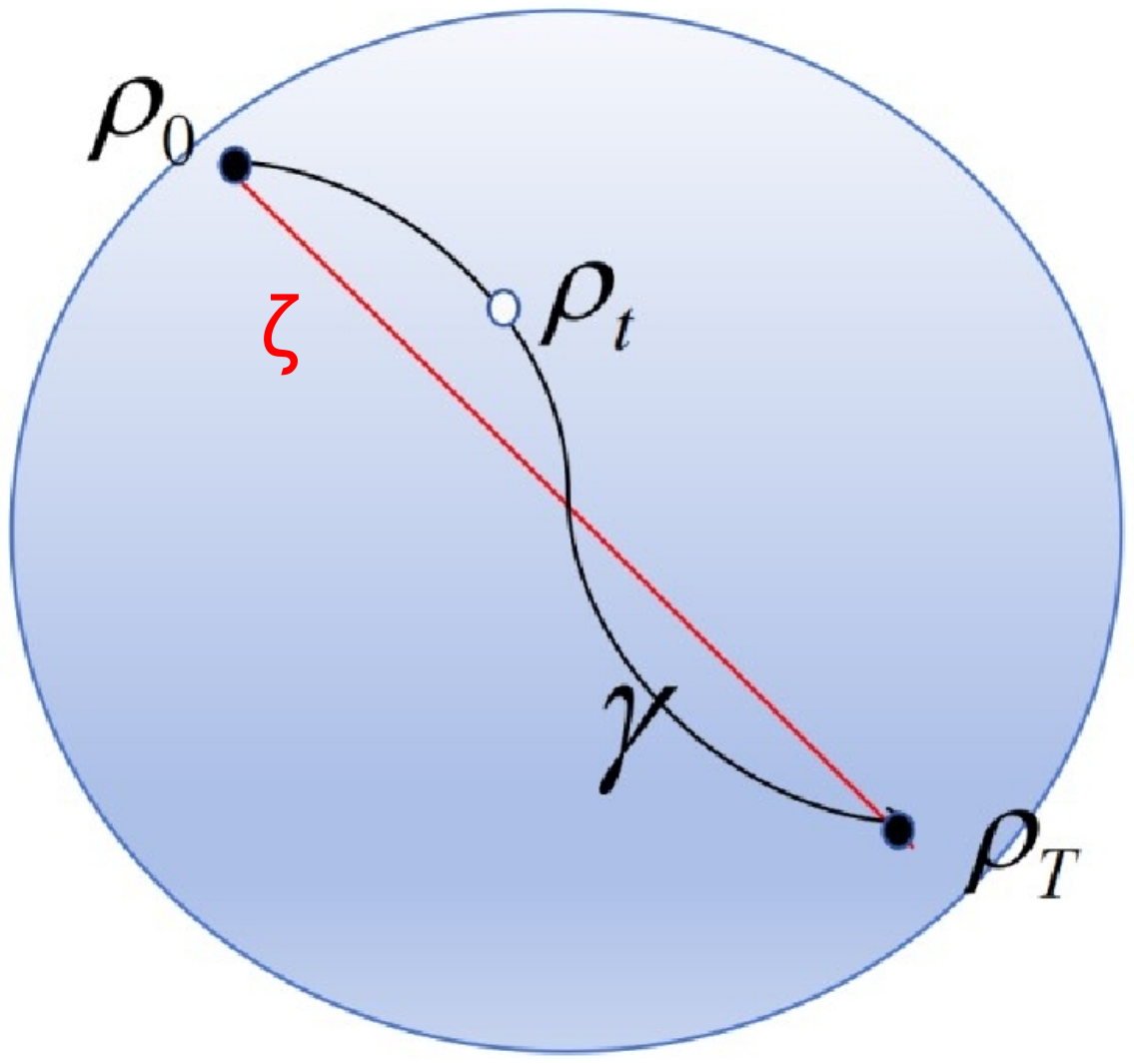}
\caption{The illustration of geometric quantum speed limits is as follows: The black curve represents the path $\gamma$ in the quantum state space, describing a general evolution from the initial state $\rho_0$ to the final state $\rho_T$, with time $t \in [0, T]$ as the parameter. Using a metric on the quantum-state space, the length of this path is denoted by $\ell_\gamma(\rho_0, \rho_T)$. The red curve $\zeta$ corresponds to the geodesic connecting $\rho_0$ and $\rho_T$, and its length is given by $\mathcal{L}(\rho_0, \rho_T)$.}
\label{geodesic}
\end{figure}

The QSL arises from the fundamental fact that among all possible physical evolutionary paths $\ell_\gamma\left(\rho_0, \rho_T\right)$ from the initial state $\rho_0$ to the final state $\rho_T$, there exit the shortest geodesic paths $\mathcal{L}\left(\rho_0, \rho_T\right) \leq \ell_\gamma\left(\rho_0, \rho_T\right)$ for arbitrary physical evolution path $\gamma$ \cite{JMP451787}. Therefore, the lower bound of the QSL is saturated when the actual evolution path of the system coincides with the geodesic $\zeta$, as shown in Fig. \ref{geodesic}.


QSLs are deeply connected to the geometric characteristics of quantum states, encompassing Fisher information, quantum state distances \cite{PRA110042425,NewJPhys24065001}, and the Bures angle \cite{SciRep105500}. They have been extensively analyzed by examining the rate of change of various physical quantities, including the expectation values of observables \cite{PRA106042436}, autocorrelation functions \cite{Quantum6884}, and different quantum properties such as quantumness \cite{srep38149}, coherence \cite{NewJPhys.24.065003,PRA110042425}, entanglement \cite{PhysRevA.104.032417}, and correlations \cite{PhysRevA.107.052419}. Additionally, QSLs have been explored in relation to quantum resources \cite{NewJPhys24065001} and diverse information-theoretic measures \cite{NewJPhys.24.065003,PhysRevE.103.032105}.

The complex numbers are indispensable in characterizing quantum states and their dynamic behaviors \cite{PhysRevA.87.052106}. The quantum imaginarity has emerged as a key concept in quantum informatics, capturing the role played by the imaginary part of quantum states or operations in quantum mechanics. The significance of quantum imaginarity is undeniable in many quantum phenomena and quantum information processing tasks \cite{JPA51414009}.
A growing body of research and experimental findings has emphasized the indispensable role of complex numbers in quantum mechanics \cite{panjw,liangxb}. In response, scientists have established a resource-theoretic approach to understanding imaginarity and have proposed multiple measures to quantify it \cite{JPA51414009,PhysRevLett.126.090401}.


The importance of the imaginary part in quantum mechanics has garnered significant attention \cite{liangxb,Quantum4359} and has been thoroughly confirmed through a series of experimental studies \cite{PhysRevLett.128.040403}. It has been demonstrated that imaginarity carries significant operational implications \cite{PhysRevLett.126.090401}, playing a crucial role in tasks such as state discrimination \cite{PhysRevA.65.050305,CommunPhys7171}, quantum hiding and masking \cite{PhysRevResearch3033176}, machine learning applications \cite{PhysRevResearch5013146}, and multiparameter quantum metrology \cite{Quantum6665}.

Motivated by the operational roles of imaginarity outlined above, it is natural to ask how fundamental dynamical speed limits depend on this resource. Complementary to our imaginarity-based formulation, the geometric and phase--based QSLs have been established. Sun \emph{et al.} quantify a unified QSL via the changing rate of the system's phase~\cite{SunPRL127_100404}, while Sun and Zheng obtain a distinct QSL from a gauge--invariant distance on projective Hilbert space~\cite{SunZhengPRL123_180403}. These approaches yield kinematic lower bounds on the minimum evolution time that are independent of any particular resource measure and apply to Hermitian (and certain non-Hermitian) dynamics. By contrast, the present work expresses QSLs in terms of imaginarity measures (trace, geometric, and relative entropy based), thereby linking dynamical speed to the creation or depletion of imaginarity as a quantum resource. The two viewpoints are complementary: phase or distance geometry constrains motion, while resource content constrains how fast that motion can be realized.
Moreover, the experimental test of real vs.\ complex quantum theory in a photonic network~\cite{LiPRL128_040402} demonstrates the operational necessity of complex numbers, which provides a direct motivation for treating imaginarity as a resource and for the imaginarity-based QSLs developed here.

In this paper, we derive QSLs for three different measures of quantum imaginarity.
We first examine the QSL based on the relative entropy imaginarity measure. We illustrate the results by dephasing and dissipative dynamics. Then we analyze the QSL based on the trace distance imaginarity measure and demonstrate its attainability. Finally, we explore the QSL based on the geometric measure of imaginarity and determine the minimum time required to generate or degrade imaginarity.
We find that the bound of QSL given by the geometric measure of imaginarity is attained when the initial state is a maximally imaginary state. We provide a condition under which the inequality is saturated by applying a unitary transformation. For simplicity, we refer to such an  imaginarity-induced QSL as imaginarity speed limit (ISL). We further use the ISL bound based on the geometric measure of imaginarity to estimate the time required for stochastic-approximate transformations. This paper is organized as follows. In Sec. \ref{sect2}, we derive the QSL bounds based on imaginarity measures and examine their attainability and tightness. We conclude with a summary and discussion in Sec. \ref{sect3}.

\section{Quantum speed limit based on imaginarity measures}\label{sect2}
In the theory of imaginarity, an orthonormal basis $\{|j\rangle\}_{j=1}^d$  of $\mathcal{H}$ is fixed, i.e., analogous to the resource theory of coherence, the imaginarity  also depends on the reference basis.
In the following, we also make use of the fact that any $d$-dimensional quantum state $\rho$ can be decomposed as~\cite{JPA51414009}
\begin{equation}\label{rhoReIm}
\rho = \text{Re}(\rho) + i\,\text{Im}(\rho),
\end{equation}
where
\begin{equation}
\text{Re}(\rho) = \frac{\rho + \bar{\rho}}{2} = \frac{\rho + \rho^T}{2}=\sum_{j, k=1}^d\left(\operatorname{Re} \rho_{j k}\right)\ket{j}\bra{k}
\end{equation}
is a real quantum state and
\begin{equation}
\text{Im}(\rho) = \frac{\rho - \bar{\rho}}{2i} = \frac{\rho - \rho^T}{2i}=\sum_{j, k=1}^d\left(\operatorname{Im} \rho_{j k}\right)\ket{j}\bra{k}
\end{equation}
is a real antisymmetric matrix, where $\rho^T$ and $\bar{\rho}$ denote the transpose and the complex conjugate of $\rho$ respectively under the selected basis.

We conclude that, for any density operator $\rho$, the operator $\operatorname{Re} \rho = \frac{1}{2} \left( \rho + \rho^{T} \right)$ is itself a valid quantum state. This fact plays an essential role in the subsequent analysis and can be verified directly: (1) (Hermitian) since $\rho$ is Hermitian, we have $\rho = \rho^\dagger$ and $\rho^T = (\rho^\dagger)^T = (\rho^T)^\dagger$, where $\rho^\dagger$ denotes the complex conjugate and transpose of $\rho$. Therefore,
\[
(\operatorname{Re} \rho)^\dagger = \left( \frac{1}{2} (\rho + \rho^T) \right)^\dagger = \frac{1}{2} (\rho^\dagger + (\rho^T)^\dagger) = \frac{1}{2} (\rho + \rho^T) = \operatorname{Re} \rho;
\]
(2) (unit trace) $\operatorname{tr}(\operatorname{Re} \rho) = \frac{1}{2} \left( \operatorname{tr}(\rho) + \operatorname{tr}(\rho^{T}) \right) = \operatorname{tr}(\rho) = 1$;
(3) (positive semi-definiteness) for any unit vector $\ket{\psi} \in \mathcal{H}$, we have
$\bra{\psi} \operatorname{Re} \rho \ket{\psi} = \frac{1}{2} \left( \bra{\psi} \rho \ket{\psi} + \bra{\psi} \rho^{T} \ket{\psi} \right)\geq 0$ as $\rho$ and $\rho^{T}$ are positive semi-definite.

Next, we recall the resource theory of imaginarity. Two key ingredients in a framework of resource theory are: the set of free states and the set of free operations~\cite{ChitambarGour2019}. Denote $\mathcal{D}(\mathcal{H})$ the convex set of all density operators $\rho$ acting on a Hilbert space $\mathcal{H}$. The free states in the resource theory of imaginarity~\cite{SciChinaPhys66280312} are real states defined by
\begin{equation*}
\mathscr{R} = \left\{ \rho \in \mathcal{D}(\mathcal{H}) : \langle j | \rho | k \rangle \in \mathbb{R}, \quad j,k = 1,2,\dots, d \right\}
\end{equation*}
with respect to a chosen basis $\{|i\rangle\}_{i=1}^d$. The set $\mathscr{R}$ consists of all density operators $\rho$ satisfying any of the following equivalent conditions: $\rho = \rho^T$, $\rho = \bar{\rho}$, $\rho = \text{Re}(\rho)$ or $\text{Im}(\rho) = 0$. Correspondingly, the free operations are the \textit{real operations} given by quantum channels $\Phi(\rho) = \sum_j K_j \rho K_j^\dagger$, where each Kraus operator $K_j$ has only real matrix entries, i.e., $\langle m | K_j | n \rangle \in \mathbb{R}$.

A nonnegative function $\mathscr{M}$ on quantum states is called an \textit{imaginarity measure} if $\mathscr{M}$ satisfies the following conditions (M1) -- (M4)~\cite{JPA51414009,QuantumInfProcess20383,PhysRevA.103.032401}:
\begin{itemize}
    \item[\textbf{(M1)}] \textbf{Non--negativity:} $\mathscr{M}(\rho) \geq 0$, and $\mathscr{M}(\rho) = 0$ for any real state $\rho$.
    \item[\textbf{(M2)}] \textbf{Monotonicity:} $\mathscr{M}[\Phi(\rho)] \leq\mathscr{M}(\rho)$ whenever $\Phi$ is a real operation.
    \item[\textbf{(M3)}] \textbf{Probabilistic monotonicity:}
    \[
    \sum_j p_j\mathscr{M}(\rho_j) \leq\mathscr{M}(\rho),
    \]
    where $p_j = \operatorname{tr}(K_j \rho K_j^\dagger)$, $\rho_j = \frac{1}{p_j} K_j \rho K_j^\dagger$, for all $\{K_j\}$ with $\sum_j K_j^\dagger K_j = I$ and $K_j$ are real operators.
    \item[\textbf{(M4)}] \textbf{Convexity:}
    \[
   \mathscr{M} \left( \sum_j p_j \rho_j \right) \leq \sum_j p_j\mathscr{M}(\rho_j),
    \]
    for any ensemble $\{p_j, \rho_j\}$.
\end{itemize}
Note that (M3) and (M4) together imply (M2).
Property (M3) implies that all imaginarity measures are invariant under real orthogonal transformations~\cite{PhysRevA.103.032401}, i.e., $\mathscr{M}(O \rho O^{\top}) =  \mathscr{M}(\rho).$
In Ref.~\cite{QuantumInfProcess20383}, the authors considered the condition (M5) below,
\begin{itemize}
    \item[\textbf{(M5)}] \textbf{Additivity for direct sum states:}
    $$
    \mathscr{M}(p \rho_1 \oplus (1-p) \rho_2) = p \mathscr{M}(\rho_1) + (1-p) \mathscr{M}(\rho_2).
   $$
\end{itemize}
It is shown that (M3) and (M4) are equivalent to (M2) and (M5)~\cite{QuantumInfProcess20383}. $\mathscr{M}$ is called an \textit{imaginarity monotone} if it satisfies all the conditions but (M4), similar to the entanglement monotone.

While imaginarity has been recognized as a quantum resource with applications in computation and metrology, its dynamical behavior remains less explored compared to other well-studied resources such as coherence and entanglement. To bridge this gap, we aim to investigate how imaginarity evolves under various dynamical scenarios. In particular, we examine its typical behavior in simple systems undergoing unitary evolution, dephasing and dissipation, so as to illustrate the motivation on the study of speed of evolution.

Existing QSLs are predominantly geometric or energy-based, providing kinematic lower bounds on the minimum evolution time. In contrast, many quantum information tasks hinge on the specifically complex (imaginary) component of quantum states and operations. Motivated by the experimental necessity of complex numbers in quantum theory~\cite{LiPRL128_040402} and by the resource theory of imaginarity,
we formulate resource-dependent QSLs in terms of imaginarity measures (trace, geometric, and relative entropy). Our bounds quantify how fast imaginarity can be generated or erased under open dynamics, yielding closed form limits for dephasing and dissipative models
(Eqs.~\eqref{dephasing_rhot},~\eqref{dissipativestate})
and an operational time $t_\varepsilon$ to reach a target threshold $\mathscr{M}_r\le\varepsilon$. This complements recent phase distance and gauge distance  QSLs~\cite{SunPRL127_100404,SunZhengPRL123_180403}: while those capture geometric constraints on motion, our ISLs reveal how the resource content itself constrains dynamical speed. Practically, one may
take the conservative bound $T \ge\max\{T_{\mathrm{phase}},T_{\mathrm{gauge}},T_{\mathrm{ISL}}\}$.

Let $\rho$ represent a density matrix in a $d$-dimensional Hilbert space $\mathcal{H}$. The evolution of a quantum system under a specified dynamical process is governed by the master equation \cite{PhysRev121920}
\begin{equation}\label{mathcalLrho}
\dot{\rho}_t=\frac{\mathrm{d} \rho_t}{\mathrm{~d} t}=\mathcal{L}_t\left(\rho_t\right),
\end{equation}
where $\rho_t$ denotes the system's density matrix at time $t$, and $\mathcal{L}_t$ refers to the corresponding Liouvillian super-operator \cite{ARivas2012}.

For closed quantum systems evolving under a Hermitian Hamiltonian \( H \), the state evolves unitarily as \( \rho(t) = e^{-iHt} \rho(0) e^{iHt} \). Since imaginarity is basis-dependent, its evolution depends on whether \( H \) commutes with the projectors defining the real subspace.

Here we first recall a measure of imaginarity that will be used in the following example--the trace distance imaginarity \cite{JPA51414009,QuantumInfProcess201}. It is defined as
\begin{equation}\label{definetracedistance}
\mathscr{M}_{tr}(\rho)=\min _{\sigma \in \mathscr{R}}\|\rho-\sigma\|_1=\frac{1}{2}\operatorname{tr}\left|\rho-\rho^T\right| =\left\|\operatorname{Im}\rho\right\|_1,
\end{equation}
where $\|A\|_{1}= \operatorname{tr}|A|=\operatorname{tr}[\sqrt{ A^{\dagger}A}]$. As proved in Ref.~\cite{JPA51414009}, the trace distance imaginarity achieves its minimum at the unique closest real state $\sigma = \operatorname{Re}(\rho) = \frac{1}{2}(\rho + \rho^T).$  This measure captures the amount of imaginarity by quantifying the distance between a quantum state and the set of real states.

As an illustrative example, consider a qubit system with Hamiltonian \( H = \omega \sigma_x \) and initial real state \( |\phi_0\rangle = |0\rangle \) in the computational basis, where $\sigma_x$ is the standard Pauli matrix. The evolved state acquires the imaginary off-diagonal elements due to the noncommutativity between \( \sigma_x \) and the reference projectors. This leads to a time-dependent increase in imaginarity. Using the trace distance measure~\eqref{definetracedistance}, one finds $\mathscr{M}_{\mathrm{tr}}(\rho_t) = \| \operatorname{Im}(\rho_t) \|_1 = |\sin(2\omega t)|$. This shows that imaginarity does not increase monotonically, but oscillates due to the interplay between the system Hamiltonian and the structure of the initial state.

If the Hamiltonian is diagonal in the reference basis, the imaginarity remains unchanged. For instance, with \( H = \omega \sigma_z \) and \( |\phi_0\rangle = |0\rangle \), we have \( |\phi_t\rangle = e^{-i\omega t} |0\rangle \) and \( \mathscr{M}_{\mathrm{tr}}(\rho_t) = 0 \) for all \( t \). In general, the imaginarity is generated only when the Hamiltonian has off-diagonal elements in the reference basis.


As a common noise process in open quantum systems, the dephasing may suppress off-diagonal elements of the density matrix. For example, in a Markovian dephasing channel described by a Lindblad equation with jump operator \( L = \sqrt{\gamma} \sigma_z \), the off-diagonal elements decay as \( \rho_{01}(t) = \rho_{01}(0) e^{-\gamma t} \). This leads to an exponential decay of imaginarity in terms of the trace distance of imaginarity.

Concerning the imaginarity under dissipation, we consider a qubit system undergoing amplitude damping, a prototypical dissipative process representing energy loss due to spontaneous emission. This type of dynamics drives the system toward a fixed steady state, which is often a real, classical-like state in the computational basis. For example, in the amplitude damping channel described by the Lindblad operator \( L = \sqrt{\gamma} |0\rangle\langle 1| \), the excited state \( |1\rangle \) decays to the ground state \( |0\rangle \) at rate \( \gamma \). As the system approaches  the  steady state \( |0\rangle\langle 0| \), the imaginarity vanishes. In this setting, the trace distance measure of imaginarity decays approximately exponentially at a rate given by the dissipative coupling, $\mathscr{M}_{\mathrm{tr}}(\rho(t)) \propto e^{-\gamma t}$.

Unlike the coherence, which may survive in certain bases under specific noise models, the imaginarity is generally more fragile, as it is directly linked to the presence of complex amplitudes. Any dissipative process that projects the system onto a real subspace or suppresses the imaginary parts of the density matrix tends to irreversibly destroy imaginarity. This fragility underscores the importance of understanding and quantifying the speed at which imaginarity degrades, particularly in open-system settings. We next investigate the QSLs based on imaginarity through three different imaginarity measures.

\subsection{ISL for relative entropy of imaginarity}
Given a fixed orthonormal basis $\{|j\rangle\}_{j=1}^d$ of the Hilbert space $\cal{H}$, the relative entropy of imaginarity for a quantum state $\rho$ is defined as \cite{SciChinaPhys66280312}
\begin{equation}
\mathscr{M}_r(\rho)=\min _{\sigma \in \mathscr{R}} S(\rho \| \sigma),\label{relativeentropy1}
\end{equation}
where $S(\rho \| \sigma) = \operatorname{tr}[\rho \ln \rho] - \operatorname{tr}[\rho\ln \sigma]$ represents the quantum relative entropy. A closed-form expression for the relative entropy of imaginarity is provided in Ref. \cite{QuantumInfProcess20383},
\begin{equation}
\mathscr{M}_r(\rho)= S(\operatorname{Re}\rho)-S(\rho),\label{relativeentropy2}
\end{equation}
where $S(\rho) = -\operatorname{tr}[\rho \ln \rho]$ denotes the von Neumann entropy, and $\operatorname{Re}\rho$ represents the real part of the density matrix $\rho$.

We first present the following Lemma \ref{lem:co-change}.

\begin{lemma}\label{lem:co-change}
 The rate of change of imaginarity $ \mathscr{M}_r(\rho_t)= S(\operatorname{Re}\rho_t)-S(\rho_t)$ is expressed as
\begin{equation}\label{eq:coherence-rate-change}
  \frac{{\rm d} }{{\rm d}t}\mathscr{M}_r(\rho_t)= -{\rm tr}[\mathcal{L}_{t}(\operatorname{Re}\rho_t) \ln \operatorname{Re}\rho_t ]+{\rm tr}[\mathcal{L}_{t}({\rho_t}) \ln \rho_t].
\end{equation}
\end{lemma}
$\mathit{Proof}$.~It is shown in Ref.~\cite{Das2018} that for any finite-dimensional quantum system, the rate of entropy change is given by
\[
\frac{{\rm d}}{{\rm d}t} S(\rho_t) = -{\rm tr}\left[ \dot{\rho}_t \ln \rho_t \right],
\]
whenever the time derivative \(\dot{\rho}_t = \mathcal{L}_t(\rho_t)\) is defined as in Eq.~(\ref{mathcalLrho}). As verified at the beginning of Sec.~\ref{sect2}, \(\operatorname{Re} \rho_t\) is itself a valid quantum state. Therefore, we have
\[
\frac{{\rm d}}{{\rm d}t} S(\operatorname{Re} \rho_t)=-{\rm tr}\left[ \dot{(\operatorname{Re} \rho_t)} \ln \operatorname{Re} \rho_t \right] = -{\rm tr}\left[ \mathcal{L}_t(\operatorname{Re} \rho_t) \ln \operatorname{Re} \rho_t \right].
\]
As a result, Eq.~(\ref{eq:coherence-rate-change}) follows immediately.$\Box$

\begin{theorem}\label{thm1}
To describe the quantum dynamics of a finite-dimensional quantum system, consider the time required to evolve from the initial relative entropy imaginary value $\mathscr{M}_r(\rho_0)$ at $t = 0$ to the final value $\mathscr{M}_r(\rho_T)$ at $t = T$. The minimum time required is constrained by the following lower bound,
\begin{equation}
T \geq T_{ISL} = \frac{|\mathscr{M}_r(\rho_T) - \mathscr{M}_r(\rho_0)|}{\Lambda_T \overline{\| \ln ( \rho_t) \|_{\text{HS}}}+\Lambda_T^{\text{Re}} \overline{\| \ln (\operatorname{Re} \rho_t) \|_{\text{HS}}}},
\label{isl}
\end{equation}
where $\|N\|_{\text{HS}} = \sqrt{\operatorname{tr}[N^\dagger N]}$ is the Hilbert-Schmidt norm of an operator $N$,
\[
\overline{\|N\|_{\text{HS}}} := \sqrt{\frac{1}{T} \int_0^T \|N\|_{\text{HS}}^2 \, dt}
\]
is the time-averaged Hilbert-Schmidt norm of any operator $N$,
$\Lambda_T^{\text{Re}} = \overline{\| \mathcal{L}_t(\operatorname{Re} \rho_t) \|_{\text{HS}}}$ represents the root mean square evolution speed of the real part of the quantum state, with $\mathcal{L}_t$ denoting the Liouvillian super-operator, and similarly,
$\Lambda_T = \overline{\| \mathcal{L}_t(\rho_t) \|_{\text{HS}}}$.
\end{theorem}
The proof and detailed derivations are provided in \textbf{Appendix A}.

The denominator of the bound in Theorem~\ref{thm1} consists of two terms: the root mean square evolution speed of the real part of the quantum state, $\Lambda_T^{\text{Re}} \, \overline{\| \ln (\operatorname{Re} \rho_t) \|_{\text{HS}}}$, and the total evolution speed of the state, $\Lambda_T \, \overline{\| \ln \rho_t \|_{\text{HS}}}$. The first term captures how rapidly and complexly the real component of the state evolves, which is intrinsically linked to the behavior of the imaginarity component through the relative entropy measure $\mathscr{M}_r$. The second term reflects the entropy like dynamical complexity of the whole system, incorporating both real and imaginary parts. Physically, this dual contribution indicates that the minimum time $T$ required to achieve a given change in $\mathscr{M}_r$ is constrained not only by the overall dynamical rate, but also by the interplay between the real and imaginary parts of the quantum state. This structure resembles traditional QSLs such as the MT or ML bounds, where the evolution time is governed by energy uncertainty and state distinguishability. Here, the bound in Theorem~\ref{thm1} is tailored to the imaginarity resource, offering a refined view of how quantum dynamics constrain the creation or degradation of imaginarity in quantum states.
In the following the time is rendered dimensionless by rescaling with the characteristic rate of the dynamics. For open--system models we use $\tau=\gamma t$ (time in unit of $\gamma^{-1}$), while for the unitary rotation example we use $\tau=\omega t$ (time in unit of $\omega^{-1}$).

Let us consider a qubit under unitary evolution governed by the Hamiltonian $H = \omega \sigma_x$. The qubit is initialized in the pure real state $\rho_0 = |0\rangle\langle0|$. At time $t$ we have
$
\rho_t = e^{-i \omega t \sigma_x} \rho_0 \, e^{i \omega t \sigma_x},
$
corresponding to a coherent rotation about the $x$-axis of the Bloch sphere. Since the initial state is real in the computational basis, the evolution generates imaginarity via the introduction of purely imaginary off-diagonal elements. At time $T = \pi/(4\omega)$, the system reaches the maximally imaginary state,
$$
\rho_T = \frac{1}{2}
\begin{pmatrix}
1 & -i \\
i & 1
\end{pmatrix},~~ with~
\operatorname{Re} \rho_T = \frac{1}{2}
\begin{pmatrix}
1 & 0 \\
0 & 1
\end{pmatrix}.
$$
Since $\rho_T$ is pure $(S(\rho_T) = 0)$ and $\operatorname{Re} \rho_T$ is maximally mixed $(S(\operatorname{Re} \rho_T) = \ln 2)$, the relative entropy of imaginarity is $\mathscr{M}_r(\rho_T)= S(\operatorname{Re}\rho)-S(\rho)= \ln 2.$


To estimate the lower bound from Theorem~\ref{thm1}, we approximate the dynamical quantities as follow:
The state $\rho_t$ is pure for all $t$. Hence, $\overline{\|\ln\rho_t\|_{HS}}=0$.
Meanwhile
\[
\operatorname{Re}\rho_t=\begin{pmatrix}
\cos^2(\omega t) & 0 \\
0 & \sin^2(\omega t)\end{pmatrix},
\]
\[\ln(\operatorname{Re}\rho_t)=\begin{pmatrix}
\ln\cos^2(\omega t) & 0 \\
0 & \ln\sin^2(\omega t)\end{pmatrix}.
\]
We have
\begin{equation*}
\overline{\|\ln(\operatorname{Re}\rho_t)\|_{HS}}
=\sqrt{\frac{4}{\pi}\!\int_0^{\pi/4}\!\big([\ln\cos^2\theta]^2+[\ln\sin^2\theta]^2\big)\,d\theta}
=:C,
\end{equation*}
where $C\simeq 3.2285.$ Since ${\cal L}_t(X)=-i[H,X]$, one finds
\begin{align}
\|{\cal L}_t(\operatorname{Re}\rho_t)\|_{\mathrm{HS}}^{2}  &=2\omega^{2}\cos^{2}(2\omega t)
\nonumber\\ \Rightarrow \Lambda_T^{\mathrm{Re}}
&=\sqrt{\frac{1}{T}\!\int_{0}^{T}\!\|{\cal L}_t(\operatorname{Re}\rho_t)\|_{\mathrm{HS}}^{2}\,dt}
=\omega,\nonumber
\end{align}
and $\Delta \mathscr{M}_r=\mathscr{M}_r(\rho_T)-\mathscr{M}_r(\rho_0)=\ln 2$ for $T=\pi/(4\omega)$.
Hence
\[
T_{\mathrm{ISL}}\ge \frac{\ln 2}{\omega\,C}\approx \frac{0.2147}{\omega},
\]
which provides a nontrivial lower bound with the correct $1/\omega$ scaling. This lower bound is smaller than the actual time $T = \pi/(4\omega)$ required to reach the maximally imaginary state, as expected. While not saturated, the bound correctly captures the scaling with $\omega$ and highlights the interplay between the real and imaginary dynamics. The example demonstrates that the bound effectively constrains the rate of imaginarity evolution up to a constant prefactor.


\textit{Remark.}
Although the ISL bound in Eq.~\eqref{isl} involves time-averaged quantities that  require some knowledge of the system dynamics, just like many formulations of QSLs, the bound  still offer valuable insights:
In the unitary rotation immediately above, all terms are obtained in closed form, yielding $\overline{\|\ln\rho_t\|_{HS}}=0$, $\overline{\|\ln(\operatorname{Re}\rho_t)\|_{HS}}=C\simeq 3.2285$, $\Lambda_T^{\operatorname{Re}}=\omega$, $\Delta\mathscr{M}_r=\ln 2$, and hence $T_{\mathrm{ISL}}=\ln 2/(\omega C)\approx 0.2147/\omega$.
This fixes a nontrivial prefactor beyond dimensional scaling and helps to characterize the fundamental constraints on how rapidly a quantum system can alter its imaginarity when the evolution is known or reasonably approximated. In addition, the decomposition into contributions from the real part and the full state helps to illuminate the role of different dynamical features. This structure could potentially assist in analyzing, benchmarking or even guiding the optimization of physical processes such as dephasing or dissipation.
For open-system dynamics, Sec.~\ref{sect2} and Apps.~D--E provide closed expressions of $\|\mathcal{L}(\rho_t)\|_{HS}$, $\|\mathcal{L}(\operatorname{Re}\rho_t)\|_{HS}$, $\|\ln\rho_t\|_{HS}$ and $\|\ln(\operatorname{Re}\rho_t)\|_{HS}$, from which we show that for any detection threshold $\varepsilon>0$, the operational time $t_\varepsilon:=\inf\{t:\mathscr{M}_r(\rho_t)\le\varepsilon\}$ is finite and, under dephasing or dissipation, scales logarithmically with $1/\varepsilon$ (Eqs.~(13),(17) and the ensuing discussion).
Taken together, the bound appears to  be  consistent with physical expectations and reflect the asymptotic behavior of imaginarity degradation in various settings.

The QSL on imaginarity is applicable to both the generation and degradation of imaginarity. Specifically, the bound presented in (\ref{isl}) determines the minimum time required for a state with imaginarity resources to transition to a real free state. If, during the evolution, the real part of $\rho_t$ in the reference basis remains unchanged, then ${\Lambda}^{\operatorname{Re}}_T$ will be zero.
The imaginarity may stay invariant in two distinct scenarios: (i) the dynamics conserve the imaginarity under the evolution, (ii) the initial state is a fixed point of the evolution.
As a result, the minimal time $T_{ISL}$ becomes zero if there is no change in imaginarity, even if the quantum state itself undergoes transformations during the evolution. To illustrate the attainability of our bound, we apply it to two quantum dynamics of particular interest: the dephasing process and the dissipative process, as discussed in \cite{DALidar,CommunMathPhys48119}.


{\textit{Dephasing dynamics.}}--Consider a two--level atom coupled to a bosonic reservoir. The Hamiltonian that governs the evolution of the total system is given by
\begin{align}
 H_{\mathrm{tot}} = \frac{1}{2} \omega_0 \sigma_z + \sum_j \omega_j b_j^\dagger b_j + \sum_j g_j \sigma_z b_j^\dagger + \mathrm{H.c.},   \label{dephase Ham}
\end{align}
where $\omega_0$ represents the atomic transition frequency, $b_j$ is the annihilation operator for the $j$-th mode of the bosonic reservoir, $\omega_j$ denotes the frequency of the $j$-th mode's harmonic oscillator, and $g_j$ is the coupling strength between the atom and the $j$-th reservoir mode.
Under standard assumptions such as weak coupling and the Born--Markov approximation, the reduced dynamics of the atomic system can be described by the master equation as follows.
In Schr\"{o}dinger representation, the master equation governing the evolution of the atomic system is expressed as \cite{breuer2002theory}
\begin{equation}\label{dephasing}
\dot{\rho}_t = -i[H_0, \rho_t] + \frac{\gamma_t}{2} \left( \sigma_z \rho_t \sigma_z - \rho_t \right),
\end{equation}
where $H_0=\frac{1}{2}\omega_0\sigma_z$ represents the free Hamiltonian of the atom, and $\gamma_t$ is the dephasing rate.
The time-dependent dephasing rate~$\gamma_t$ in \eqref{dephasing} reflects the influence of the spectral properties of the reservoir and the coupling constants~$g_j$. This mapping from the total Hamiltonian in ~\eqref{dephase Ham} to the effective master equation is a standard procedure in the theory of open quantum systems and is detailed in Ref.~\cite{breuer2002theory}. In the Markovian limit, $\gamma_t$ becomes a constant $\gamma$, whereas in the non-Markovian case, it retains explicit time dependence~\cite{Bera2021nonmarkovian,Breuer2002nonmarkovian}.

The initial state is denoted by $\rho_{0}=\op{\psi(0)}$, where the state vector $|\psi(0)\rangle$ is given by $|\psi(0)\rangle = {\cos\frac{\theta}{2}}|0\rangle +  i\sin{\frac{\theta}{2}}|1\rangle$,
with $|0\rangle$ and $|1\rangle$ representing the computational basis states, and $\theta $ being a parameter that characterizes the state. The corresponding density matrix for the initial state is
\begin{equation}\label{ini_state}
    \rho_0=\left(\begin{array}{ll}
        {\cos^2\frac{\theta}{2}} & -i{\cos\frac{\theta}{2}}\sin{\frac{\theta}{2}}\\
        i{\cos\frac{\theta}{2}}\sin{\frac{\theta}{2}} & \sin^2{\frac{\theta}{2}}
    \end{array}\right).
\end{equation}

From (\ref{dephasing}) one obtains
\begin{equation}\label{dephasing_rhot}
    \rho_t=\left(\begin{array}{ll}
       {\cos^2\frac{\theta}{2}} & -i{\cos\frac{\theta}{2}}\sin{\frac{\theta}{2}}e^{-\int_0^t dt'\gamma_{t'}-i\omega_0t}\\
       i{\cos\frac{\theta}{2}}\sin{\frac{\theta}{2}}e^{-\int_0^t dt'\gamma_{t'}+i\omega_0t} & \sin^2{\frac{\theta}{2}}
    \end{array}\right).
\end{equation}
To estimate the bound (\ref{isl}) of Theorem \ref{thm1}, we calculate the following quantities:
(i)$|\mathscr{M}_r(\rho_{t})-\mathscr{M}_r(\rho_{0})|$,
(ii)$\norm{\mathcal{L}(\rho_{t})}_{HS}^{2}$,
(iii)$\norm{\mathcal{L}(\operatorname{Re}\rho_{t})}_{HS}^{2}$,
(iv)$\norm{\ln{\operatorname{Re}\rho_{t}}}_{HS}^{2}$,
(v)$\norm{\ln{\rho_{t}}}_{HS}^{2}$.
All closed-form expressions corresponding to these quantities can be found in \textbf{Appendix D}
(Eqs.~(\ref{eq:D1})--(\ref{eq:D5})).

In this work, we compute the relative entropy of imaginarity in the computational basis $\{\ket{0},\ket{1}\}$. During the dephasing process, the diagonal elements of $\rho_t$ in the reference basis remain unchanged. As a result, ${\Lambda}^{\operatorname{Re}}_T$ vanishes, which reflects the property that the real part of $\rho_t$ remains invariant under the dephasing dynamics in the computational basis. Therefore, the entire change in imaginarity arises from the decay of the imaginary component. Hereafter we work in the interaction (rotating) picture with respect to the Hamiltonian, i.e., we set $\omega_0=0$ so that the factors $e^{\pm i\omega_0 t}$ are removed. With our choice of phase, the off--diagonal element remains purely imaginary and equals to $\pm i\,\cos(\tfrac{\theta}{2})\sin(\tfrac{\theta}{2})\,e^{-\int_0^t\gamma_{t'}dt'}$. Hence $\operatorname{Re}\rho_t=\mathrm{diag}(\cos^2\!\tfrac{\theta}{2},\,\sin^2\!\tfrac{\theta}{2})$, whose spectrum is independent of $\gamma_t$.

\begin{figure}
    \centering
    \includegraphics[width=0.95\linewidth]{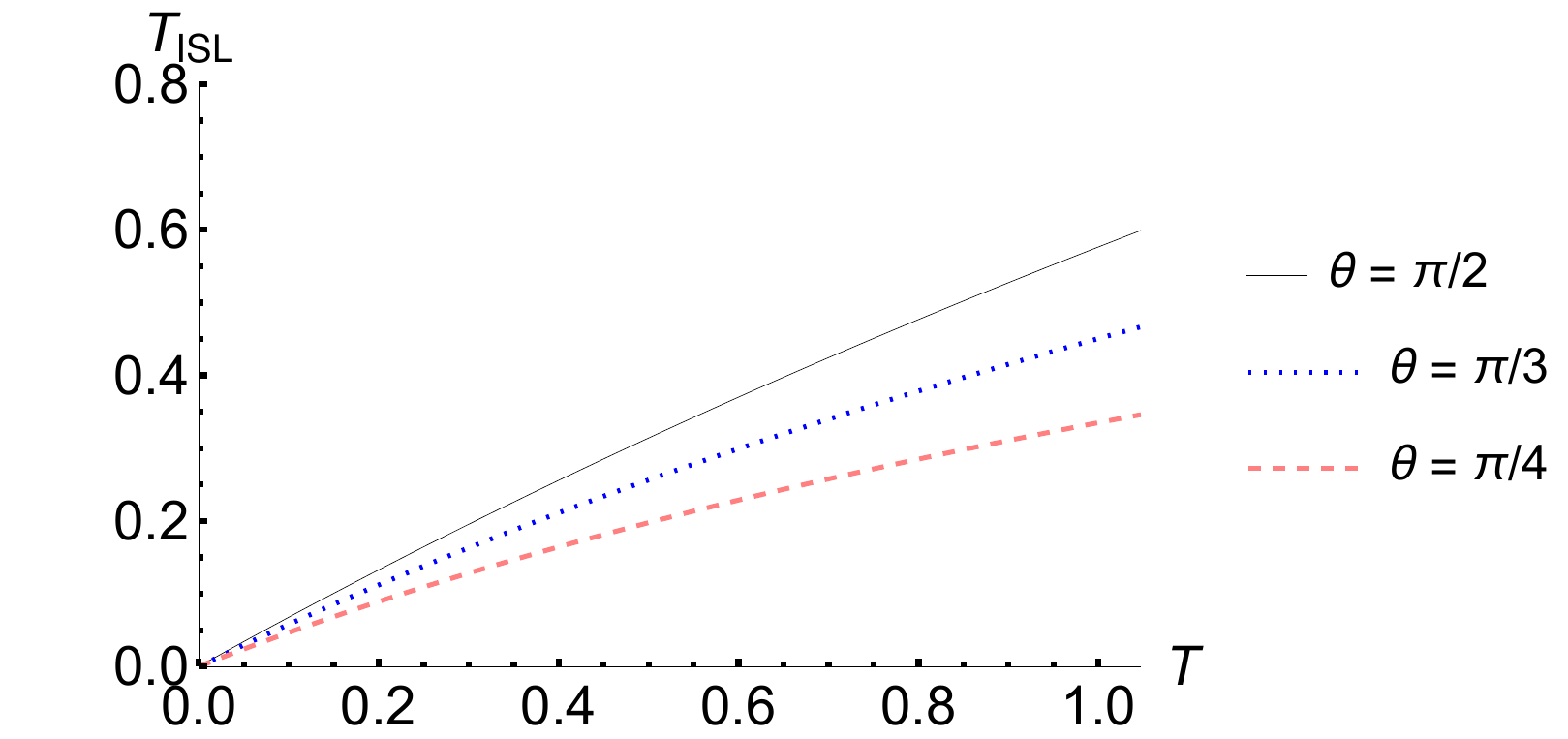}
    \caption{The evolution time $T$ under the dephasing channel is compared to the speed limit time $T_{\mathrm{ISL}}$ [as defined in \eqref{isl}], and the dynamics comes from the
master equation specified by~\eqref{dephasing}, with the parameters set as $\gamma_t = 2$ and $\omega_0 = 0$. The black solid, blue dot and pink dash lines represent the cases of $\theta=\frac{\pi}{2}$, $\frac{\pi}{3}$ and $\frac{\pi}{4}$ of the initial state, respectively.  All times are in units of $(\gamma_t)^{-1}$ (the horizontal axis shows the dimensionless time $\gamma_t t$).
   }
    \label{dephasing_fig1}
\end{figure}

In Fig.~\ref{dephasing_fig1}, we present a plot of $T_{ISL}$  as a function of $T$  over the interval $T$ $\in$ $[0, \frac{\pi}{3}]$.
It is observed that, during the dephasing process, the maximally imaginary state (MIS) \cite{JPA51414009}, given by
\begin{equation}\label{maximallystate}
|\hat{+}\rangle=\frac{1}{\sqrt{2}}(|0\rangle+i|1\rangle),
\end{equation}
shows a higher speed limit time for the reduction of imaginarity when compared to other states.
Specifically, for clarity we have chosen a constant dephasing rate $\gamma_t = 2$. This allows us to focus on the effect of the initial state's imaginarity, parameterized by $\theta$, on the speed limit time. As $\theta$ increases, the initial state becomes closer to the MIS, leading to a slower decay of the imaginarity and correspondingly larger $T_{\mathrm{ISL}}$. This is in consistent with the fact that MIS carries the highest initial resource.

Although Fig.~\ref{dephasing_fig1} uses a constant dephasing rate for simplicity, our general framework permits arbitrary time-dependent rates $\gamma_t$, which can model non-Markovian environments. In such cases, one might observe rich behavior such as the revival of imaginarity, oscillations or non-monotonic decay in $\mathscr{M}_r(\rho_t)$. These features would affect the denominator in Eq.~\eqref{isl}, thereby altering the bound $T_{\mathrm{ISL}}$ in nontrivial ways. This provides new ideas for future research on the evolution of virtual time resources and QSL.

From \eqref{dephasing_rhot}, the dephasing time for the state $\rho_t$ is infinite. In the pure dephasing model the off--diagonal term decays as $\rho_{01}(t)=i\,c\,e^{-\Gamma(t)}$ with $\Gamma(t)=\int_0^t dt'\gamma_{t'}$ and $c>0$. Since $e^{-\Gamma(t)}>0$ for any finite $t$, the imaginarity of the state--quantified here by the relative entropy of imaginarity--decreases exponentially but vanishes only in the limit $t\to\infty$. In other words, $\mathscr{M}_r(\rho_t)$ approaches to zero monotonically but never vanishes at any finite time. The ``exponential decay'' refers to the rate of approaching zero, while the infinite dephasing time means that the exact value $\mathscr{M}_r=0$ is attained only asymptotically. Our ISL bound $T_{\mathrm{ISL}}$ captures this behavior, confirming that the complete loss of imaginarity requires infinite time under pure dephasing.


Mathematically, \(\mathscr{M}_r(\rho_t)\) vanishes only asymptotically. Operationally, for any finite detection threshold \(\varepsilon>0\), there is a finite time \(t_\varepsilon\) such that \(\mathscr{M}_r(\rho_{t_\varepsilon})\le\varepsilon\) (for pure dephasing, \(t_\varepsilon\) scales as \(\ln(1/\varepsilon)\)). Our ISL bound applies directly to this setting by replacing the target $\mathscr{M}_r=0$ with $\mathscr{M}_r\le\varepsilon$, i.e., $\mathscr{M}_r(\rho_0)-\varepsilon$ vanishes at a finite time $t_\varepsilon$.

{\textit{Dissipative dynamics.}}--Suppose a two-level atom interacts with a leaky single-mode cavity \cite{breuer2002theory, PhysRevA.55.2290, PhysRevLett.111.010402}. The Hamiltonian of this system reads
\begin{equation}\label{dissipative Ham}
H_{\mathrm{tot}}=\frac{1}{2}\omega_0\sigma_z+\sum_j\omega_jb_j^\dagger b_j+\sum_j g_j\sigma_+b_j+\mathrm{H.c.}.
\end{equation}
Assume that the cavity is initially in the vacuum state and acts as a memoryless (Markovian) reservoir. The corresponding master equation is derived by using the Born-Markov and rotating wave approximations, and tracing out the cavity mode degrees of freedom~\cite{DALidar,CommunMathPhys48119}, the reduced dynamics of the atom can be described by the master equation as follows,
\begin{equation}
\dot\rho_t=\frac{\gamma_t}{2}\left(\sigma_-\rho_t\sigma_+-\frac{1}{2}
\left\{\sigma_+\sigma_-,\rho_t\right\}\right),\label{equa}
\end{equation}
here $\rho_t$ represents the density matrix of the system under consideration, while $\gamma_t$ denotes the time-dependent decay rate.
The decay rate \(\gamma_t\) in Eq.~\eqref{equa} is related to the atom--cavity coupling strengths \(g_j\) and the spectral density of the cavity modes evaluated at the atomic transition frequency \(\omega_0\). In the Markovian regime, \(\gamma_t\) is typically approximated as a constant, with $\gamma \propto 2\pi \sum_j |g_j|^2 \delta(\omega_j - \omega_0)$ as commonly found in open quantum systems theory~\cite{breuer2002theory, PhysRevA.55.2290, PhysRevLett.111.010402}.
The operators $\sigma_{+}=|0\rangle\langle 1|$ and $\sigma_{-}=|1\rangle\langle 0|$ are the atomic raising and lowering operators, respectively.

Let the initial state of the system be given by (\ref{ini_state}). One has from  (\ref{equa})
\begin{equation}
    \rho_t=\left(\begin{matrix}
        e^{-\int_0^t dt'{\frac{\gamma_t'}{2}}}\cos^2 {\frac{\theta}{2}} &-i e^{-\int_0^t dt'{\frac{\gamma_t'}{4}}} \sin {\frac{\theta}{2}} \cos {\frac{\theta}{2}}\\
       ie^{-\int_0^t dt'{\frac{\gamma_t'}{4}}} \sin {\frac{\theta}{2}} \cos {\frac{\theta}{2}} & 1-e^{-\int_0^t dt'{\frac{\gamma_t'}{2}}}\cos^2 {\frac{\theta}{2}}
    \end{matrix}\right).\label{dissipativestate}
\end{equation}

To estimate the bound in Eq.~(9) of Theorem~1, we evaluate the following quantities:
(i) $\mathscr{M}_r(\rho_{0})$,
(ii) $\mathscr{M}_r(\rho_{t})$,
(iii) $\|\mathcal{L}(\rho_t)\|_{HS}^2$,
(iv) $\|\mathcal{L}(\operatorname{Re}\rho_t)\|_{HS}^2$,
(v) $\|\ln(\operatorname{Re}\rho_t)\|_{HS}^2$, and
(vi) $\|\ln \rho_t\|_{HS}^2$.
All corresponding closed-form expressions  are provided in Appendix~E (Eqs.~(\ref{eq:E1})--(\ref{eq:E6})).

In Fig.~\ref{dissipative _fig2}, we present a plot of \( T_{ISL} \) as a function of $T$ over the interval $T$ $\in$ $[0, \frac{\pi}{3}]$ for the dissipative dynamics, with a constant decay rate $\gamma_t=2$ and $\theta\in\{\frac{\pi}{2},\frac{\pi}{3},\frac{\pi}{4}\}$. It is evident that the MIS, as described in \eqref{maximallystate}, demonstrates a higher speed limit time for the reduction of imaginarity compared to other states throughout the dissipative process. Additionally, from \eqref{dissipativestate}, we find that the dissipation time for the state $\rho_t$ is infinite, indicating that the dissipative process requires an infinite amount of time to completely remove the imaginarity from the quantum system.

We emphasize that the dissipative model governed by  Eq.~\eqref{equa} offers a rich structure for understanding the temporal evolution of imaginarity. Specifically, the quantum state undergoes energy dissipation toward the ground state, leading to an eventual reduction of coherence and imaginarity. However, as shown in Eq.~\eqref{dissipativestate}, this process occurs asymptotically, meaning that the complete elimination of imaginarity requires infinite time. This behavior is reflected in the quantum speed limit $T_{\mathrm{ISL}}$, which grows with time but never sharply drops to zero. It characterizes the fact that the system retains a small but non-negligible amount of imaginarity throughout the evolution.

Furthermore, while Fig.~\ref{dissipative _fig2} is plotted under the assumption of a constant decay rate $\gamma_t = 2$, the framework is general and allows for analyzing the effect of time-dependent decay. In non-Markovian regimes, where $\gamma_t$ can be negative or oscillatory due to reservoir memory effects~\cite{Bera2021nonmarkovian,Breuer2002nonmarkovian}, the evolution of imaginarity may exhibit temporary revivals, resulting in non--monotonic behavior in $T_{\mathrm{ISL}}$.
This suggests that the ISL bound may be sensitive to non--Markovian effects in this model.



In the dissipative case as well, the imaginarity does not vanish at a finite time but decays asymptotically. As seen in Eq.~(\ref{dissipativestate}), the populations relax exponentially toward the ground state and the imaginary coherence components decay exponentially (e.g., $\rho_{01}(t)\propto e^{-\Gamma(t)}$ with $\Gamma(t)=\int_0^t\gamma_{t'}dt'$).
Substituting the solution (Eq.~(\ref{dissipativestate})) into $\mathscr{M}_r(\rho_t)=S(\operatorname{Re}\rho_t)-S(\rho_t)$
yields a monotonically decreasing function with $\mathscr{M}_r(\rho_t)>0$ at every finite $t$ and
$\mathscr{M}_r(\rho_t)\to 0$ only as $t\to\infty$. Here, ``exponential decay'' specifies the rate of approaching zero, whereas the statement that the dissipation time is infinite means that the exact erasure of imaginarity is achieved only asymptotically. This fact aligns with the general behavior of resource measures under irreversible dynamics. The ISL bound in this case is
consistent with this interpretation, indicating that although imaginarity can decrease rapidly, its complete vanishing is fundamentally unachievable within finite time.

Mathematically, \(\mathscr{M}_r(\rho_t)\) vanishes only asymptotically under dissipative dynamics (Eqs.~\eqref{equa}--\eqref{dissipativestate}). Operationally, for any finite detection threshold \(\varepsilon>0\) there exists a finite time \(t_\varepsilon\) such that \(\mathscr{M}_r(\rho_{t_\varepsilon})\le\varepsilon\) (for constant \(\gamma\), \(t_\varepsilon\propto \ln(1/\varepsilon)\)). Our ISL bound can be used for this operational target by replacing $\mathscr{M}_r=0$ with $\mathscr{M}_r\le\varepsilon$, i.e., taking $\Delta\mathscr{M}_r \mapsto \mathscr{M}_r(\rho_0)-\varepsilon$.

\begin{figure}
\centering
\includegraphics[width=0.95\linewidth]{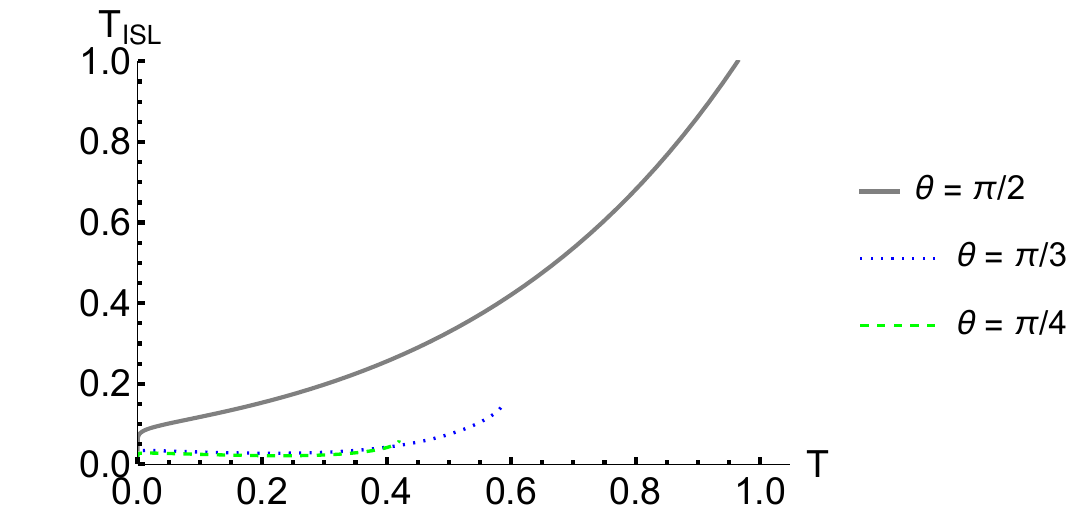}
\caption{The evolution time $T$ under the dissipative channel is compared with the speed limit time $T_{\mathrm{ISL}}$ [as defined in \eqref{isl}] and the dynamics comes from the
master equation specified by~\eqref{equa} with $\gamma_t = 2$. The black solid, blue dot and green dash lines represent the cases of $\theta=\frac{\pi}{2}, \frac{\pi}{3}$ and $\frac{\pi}{4}$ of the initial state, respectively. All times are in units of $(\gamma_t)^{-1}$ (the horizontal axis shows the dimensionless time $\gamma_t t$).
}
\label{dissipative _fig2}
\end{figure}

\subsection{ISL for trace distance imaginarity}

Using the trace distance imaginarity~\eqref{definetracedistance}, we are able to characterize the dynamical constraints on imaginarity under general physical evolutions.  We have the following theorem on ISL, which bounds the rate at which imaginarity changes over time.

\begin{theorem}\label{thm2}
For the dynamical evolution of a system from the initial state $\rho_0$ to the final state $\rho_T$, the time $T$ required for the change in imaginarity $\Delta_I$ is bounded from below by
\begin{equation}\label{traceisl}
   T \geq
   T_{\mathrm{ISL}}=\frac{\left\vert\Delta_I\right\vert}{\Lambda_{\mathrm{tr}}(T)},
\end{equation}
where $\Delta_I= \mathscr{M}_{tr}(\rho_T)-\mathscr{M}_{tr}(\rho_0)$, $\Lambda_{\mathrm{tr}}(T)
= \frac{1}{T} \int_0^T \operatorname{tr} \left| \operatorname{Im} \dot{\rho}_t \right| \, dt
= \frac{1}{T} \int_0^T \frac{1}{2} \operatorname{tr} \left| \dot{\rho}_t-\dot{\rho}_t^{T} \right| \, dt$.
\end{theorem}

$\mathit{Proof}$.  Using the fact that $\frac{d\|A\|}{dt} \leq \| \frac{dA}{dt} \|$ for any matrix norm $\|\cdot\|$ and for any matrix $A$, we have,
\begin{align}
\frac{1}{T}\int_0^T\left|\frac{\mathrm{~d}}{\mathrm{~d} t} \mathscr{M}_{tr}(\rho_t)\right| \mathrm{d} t \leq \frac{1}{T}\int_0^T    \operatorname{tr}\left| \operatorname{Im} \dot{\rho}_t \right| \mathrm{d} t. \label{trace7}
\end{align}
The inequality (\ref{traceisl}) is a direct result of (\ref{trace7}).  $\Box$

The upper bound presented in Theorem \ref{thm2} indicates the minimum time necessary for any physical process to produce a detectable change in the imaginarity of a quantum system.
We now demonstrate that the aforementioned bound is saturated when the system evolves along the geodesic. First, consider an initial state \(\rho_0\), and define the geodesic ``dephasing'' evolution as
\[
\rho(t) = \operatorname{Re} \rho_0 + i\, p(t)\, \operatorname{Im} \rho_0,
\]
where \(p(t) = 1 - t\) and \(0 \leq t \leq 1\). This evolution lasts for a total time \(T = 1\). It is straightforward to observe that: (i) the real part of the state remains constant throughout the evolution; (ii) the final state is real; and (iii) the closest real state at all times is $\sigma_t = \operatorname{Re} \rho_t = \operatorname{Re} \rho_0$.
It is clear that \(\rho(t)\) is a valid quantum state for all \(t\), since it is a convex combination of the two states \(\rho_0\) and \(\operatorname{Re} \rho_0\). Recall that \(\operatorname{Re}(\rho_t)\) is itself a valid density matrix, as shown at the beginning of Sec.~\ref{sect2}. With this definition, we find
\begin{itemize}
    \item \(\mathscr{M}_{\mathrm{tr}}(\rho(0)) = \operatorname{tr} \left| \operatorname{Im} \rho_0 \right|\),
    \item \(\mathscr{M}_{\mathrm{tr}}(\rho(T)) = \operatorname{tr} \left| \operatorname{Im} \rho(T) \right| = 0\),
    \item \(\left| \operatorname{Im} \dot{\rho}_t \right| = \left| \operatorname{Im} \rho_0 \right| \quad \Rightarrow \quad \Lambda_{\mathrm{tr}} = \operatorname{tr} \left| \operatorname{Im} \rho_0 \right|\).
\end{itemize}
Substituting these into Eq.~(\ref{traceisl}), we obtain \(T_{\mathrm{ISL}} = 1 = T\), thereby demonstrating that the ISL bound is saturated by this evolution.

\textit{Remark.}
Since $\mathscr M_{\mathrm{tr}}(\rho_t)=\|\operatorname{Im}\rho_t\|_1$, the pointwise inequality
\[
\Bigl|\tfrac{d}{dt}\,\mathscr M_{\mathrm{tr}}(\rho_t)\Bigr|
\le \|\operatorname{Im}\dot{\rho}_t\|_1
=\mathrm{tr}\,|\operatorname{Im}\dot{\rho}_t|
\]
holds for all $t$, where $\mathscr M_{\mathrm{tr}}$ is differentiable. Hence the integral bound in Eq.~\eqref{traceisl} follows by averaging. The lower bound is saturated whenever equality holds almost everywhere in time. A simple sufficient condition is a ``radial'' evolution in the imaginary sector, $\operatorname{Im}\rho_t=p(t)\,\operatorname{Im}\rho_0$, $p(0)=1$ and $p(t)$ is monotonic. Then $\mathscr M_{\mathrm{tr}}(\rho_t)=|p(t)|\,\|\operatorname{Im}\rho_0\|_1$ and
$\|\operatorname{Im}\dot{\rho}_t\|_1=|p'(t)|\,\|\operatorname{Im}\rho_0\|_1$, so that
\[
\Bigl|\tfrac{d}{dt}\,\mathscr M_{\mathrm{tr}}(\rho_t)\Bigr|
=\|\operatorname{Im}\dot{\rho}_t\|_1
\]
almost everywhere in time, and hence $T_{\mathrm{ISL}}=T$.
This covers pure dephasing Lindblad dynamics with an arbitrary (possibly time-dependent) rate $\gamma_t$ (where $p(t)=e^{-\int_0^t\gamma_{t'}dt'}$) as well as the linear geodesic $p(t)=1-t$ used for illustration. More generally, the bound serves as a tight approximation whenever $\operatorname{Im}\rho_t\approx u(t)X$ with a fixed operator $X$ (the direction varies slowly), so that the average slack
$
\frac{1}{T}\!\int_0^T\!\left(\mathrm{tr}\,|\operatorname{Im}\dot{\rho}_t|
-\Bigl|\tfrac{d}{dt}\,\mathscr M_{\mathrm{tr}}(\rho_t)\Bigr|\right)dt
$
is small. In contrast, when the dynamics mixes different imaginary components (for example, Hamiltonian rotations that couple real and imaginary parts), the inequality is strict and the ISL becomes looser.


\subsection{ISL for geometric imaginarity}
For a pure state $|\psi\rangle$, the geometric imaginarity is defined as \cite{PhysRevA.103.032401},
\begin{equation}
\mathscr{M}_g(|\psi\rangle)=1-\max _{|\phi\rangle \in \mathscr{R}}|\langle \phi | \psi \rangle|^2,
\end{equation}
where the maximization is performed over all real pure states. For a mixed state $\rho$, the geometric imaginarity $\mathscr{M}_g(\rho)$ is given by
\begin{equation}
\mathscr{M}_g(\rho)=\min \sum_j p_j \mathscr{M}_g\left(\left|\psi_j\right\rangle\right),
\end{equation}
where the minimum is taken over all pure state decompositions of $\rho=\sum_j p_j\left|\psi_j\right\rangle\left\langle\psi_j\right|$. The geometric imaginarity can also be expressed as a distance-based measure, which is equivalent to the following definition \cite{NewJPhys12123004},
\begin{align}
\mathscr{M}_g(\rho)=1-\max _{\sigma \in \mathscr{R}} F(\rho, \sigma)=\frac{1-\sqrt{F\left(\rho, \rho^T\right)}}{2},\label{geometric1}
\end{align}
where $\sqrt{F(\rho, \sigma)}=\operatorname{tr} \sqrt{\sqrt{\rho} \sigma \sqrt{\rho}}$ denotes the square root of the fidelity between the states $\rho$ and $\sigma$.

Set $\Theta_B(\rho,\sigma)=\arccos \sqrt{F(\rho,\sigma)}$ to be the Bures angle \cite{SciRep105500}. We have
\begin{align}
    \min_{\sigma\in \mathscr{R}}\Theta_B(\rho,\sigma)
    &=\arccos\left\{\max_{\sigma\in\mathscr{R}}\sqrt{F(\rho,\sigma)}\right\}\nonumber\\
    &=\arccos\sqrt{1-(1-\max_{\sigma\in \mathscr{R}}\left[F(\rho,\sigma)\right])}\nonumber\\
    &=\arccos\sqrt{1-\mathscr{M}_g(\rho)}.\label{theta1}
\end{align}
If $\mathscr{M}_g(\rho_0)>\mathscr{M}_g(\rho_T)$, we obtain
\begin{align}\label{ineq1}
&\arccos\sqrt{1-\mathscr{M}_g(\rho_0)}-\arccos\sqrt{1-\mathscr{M}_g(\rho_T)}\nonumber\\
&= \Theta_B(\rho_0,\rho_0^\star)-\Theta_B(\rho_T,\rho_T^\star)\leq \Theta_B(\rho_0,\rho_T^\star)-\Theta_B(\rho_T,\rho_T^\star)\nonumber\\
&\leq\Theta_B(\rho_0,\rho_T)\leq\int_0^T dt\Theta_B(\rho_t,\rho_{t+dt}),
\end{align}
here the last two inequalities are derived by applying the triangle inequality to the distance function $\Theta_B$. For an infinitesimally small time interval $dt$, we obtain
\begin{align}\label{metric}
   {\Theta_B}^2(\rho_t,\rho_{t+dt})=\mathrm{tr}\left(\frac{d}{dt}\sqrt{\rho_t}\right)^2,
\end{align}
see the derivation in \textbf{Appendix A}.

Similarly, for the case where \( \mathscr{M}_g(\rho_0) < \mathscr{M}_g(\rho_T) \), we obtain
\begin{align}\label{ineq2}
&\arccos\sqrt{1-\mathscr{M}_g(\rho_T)}-\arccos\sqrt{1-\mathscr{M}_g(\rho_0)}\nonumber\\
&=\Theta_B(\rho_T,\rho_T^\star)-\Theta_B(\rho_0,\rho_0^\star)\leq\Theta_B(\rho_T,\rho_0^\star)
-\Theta_B(\rho_0,\rho_0^\star)\nonumber\\
&\leq\Theta_B(\rho_0,\rho_T)\leq\int_0^T dt\Theta_B(\rho_t,\rho_{t+dt}).
\end{align}
Summarizing (\ref{ineq1}) and (\ref{ineq2}), we derive the following theorem.

\begin{theorem}\label{thm3}
For the dynamical evolution of a system from the initial state $\rho_0$  to the final state $\rho_T$, the time $T$ needed for a change in imaginarity $\Delta_I$ is bounded below by
\begin{align}\label{geometricisl}
T\geq T_{\mathrm{ISL}}=\frac{\left\vert\Delta_I\right\vert}{\Lambda_{g}(T)},
\end{align}
where $\Delta_I= \arccos\sqrt{1-\mathscr{M}_g(\rho_T)}-\arccos\sqrt{1-\mathscr{M}_g(\rho_0)}$  and  $\Lambda_{g}(T):=\frac{1}{T} \int_0^T \mathrm{d}t\sqrt{\mathrm{tr}\left(\frac{d}{dt}\sqrt{\rho_t}\right)^2}$.
\end{theorem}

Noisy environments may significantly affect the imaginarity of quantum states. Consequently, the imaginarity observed in actual quantum state evolution often differs from that in idealized environments.
The bound in Theorem \ref{thm3} provides the minimum timescale required to produce a certain change in the imaginarity of a system under any physical process. In the following, based on this bound we further investigate the minimal time required for the generation of quantum imaginarity and the minimal time required for its degradation.


Let $\textit{M}_g(\rho_t)$ denote the corresponding imaginarity under the actual dynamic evolution, and $\mathscr{M}_g(\rho_t)$ the one corresponding to the closest real state in the given dynamics. We have for any $t \in\left[0, T\right]$,
$\textit{M}_g(\rho_t)\leq\mathscr{M}_g(\rho_t)$, where $\textit{M}_g(\rho_0)=\mathscr{M}_g(\rho_0)$ for $t=0$.

Consider the minimum time needed to generate imaginarity.  If the system starts with a real state, applying the bound in \eqref{geometricisl} we derive the following lower bound on the time needed to generate a specific amount of imaginarity,
$$
T\geq T_{\mathrm{ISL}}=\frac{\mathscr{M}_g(\rho_t)}{\Lambda_{g}(T)}
\geq \frac{\textit{M}_g(\rho_t)}{\Lambda_{g}(T)}.
$$
This bound provides the minimum time required to generate a given amount of imaginarity in a
system started with a real state for any physical process.

Next consider that the system evolves from a quantum state with nonzero imaginarity to a real final state, with a complete loss of imaginarity. We have the following lower bound on the minimum time required for such degradation, that is, the time needed to completely deplete the initial amount of imaginarity,
$$
T\geq T_{\mathrm{ISL}}=\frac{\mathscr{M}_g(\rho_0)}{\Lambda_{g}(T)}
\geq \frac{\textit{M}_g(\rho_0)}{\Lambda_{g}(T)}.
$$
This bound provides an estimation of the minimum duration during which the imaginarity survives in a quantum system. It also serves to assess the time needed for the degradation of imaginarity due to a dynamical map. Accordingly, the bound offers valuable understanding of the imaginarity dynamics in quantum systems influenced by environmental noise and dissipative processes.

Let us consider the following examples to illustrate the applications.

{\textit{Dephasing dynamics.}}--Consider a two-level atom interacting with a bosonic reservoir, with the Hamiltonian given by (\ref{dephase Ham}) and the corresponding master equation (\ref{dephasing}) in the Schr\"{o}dinger representation. Assuming the initial state of the system is described by (\ref{ini_state}), solving equation (\ref{dephasing}) yields the density matrix $\rho_t$ given by (\ref{dephasing_rhot}). To further investigate the dephasing dynamics, we numerically analyze different initial states by setting $\theta=\frac{\pi}{2}$, $\frac{\pi}{3}$ and $\frac{\pi}{4}$ for the initial state (\ref{ini_state}), respectively. We use a decay rate of $\gamma_t=2$ and a frequency difference of $\omega_0=0$. The related calculations and results are presented in Fig. \ref{dephasing_fig2}, with further details provided in \textbf{Appendix B}.
Comparing the numerical results of the ISL bound (\ref{isl}) for the relative entropy imaginarity with the ISL bound (\ref{geometricisl}) for the geometric imaginarity, Figs. \ref{dephasing_fig1} and \ref{dephasing_fig2}, it is seen that the speed limit constraints of the geometric imaginarity show preferable tightness for this model.
\begin{figure}
    \centering
    \includegraphics[width=0.95\linewidth]{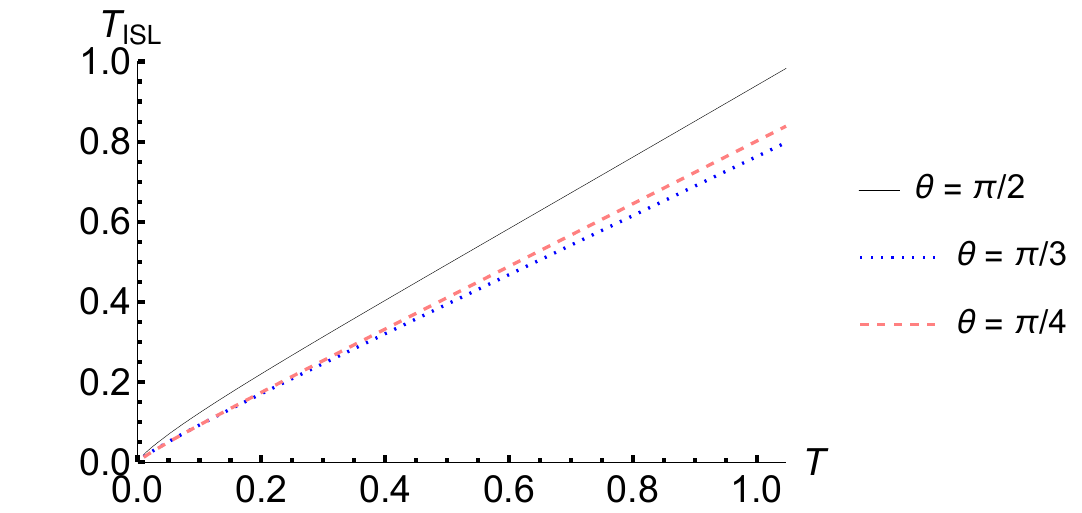}
    \caption{
    The actual evolution time $T$ under the dephasing channel is compared to the speed limit time $T_{\mathrm{ISL}}$  defined in \eqref{geometricisl} and the dynamics comes from the
master equation specified by~\eqref{dephasing}, with the parameters set as $\gamma_t = 2$ and $\omega_0 = 0$. The black solid, blue dotted, and pink dashed lines correspond to initial state parameters of $\theta = \frac{\pi}{2}$, $\theta = \frac{\pi}{3}$, and $\theta = \frac{\pi}{4}$, respectively. All times are in units of $(\gamma_t)^{-1}$ (the horizontal axis shows the dimensionless time $\gamma_t t$).
   }
    \label{dephasing_fig2}
\end{figure}


Specifically, we find that the speed limit constraint for the geometric measure of imaginarity reaches its maximum when the initial state is a MIS, as given in  \eqref{maximallystate}.
We can impose the above dephasing process under the Markovian regime in this representation and obtain similar results. In this sense, we can conclude that for any initial state, one can always construct proper dephasing dynamics such that the system's imaginarity is degraded along the geodesics of the state evolution, because any arbitrary initial state can be obtained by applying a unitary operation to the MIS. Therefore, the ISL bound in Eq.~\eqref{geometricisl}, which is based on the geometric measure of imaginarity, is indeed achievable.

\begin{figure}
    \centering
    \includegraphics[width=1\linewidth]{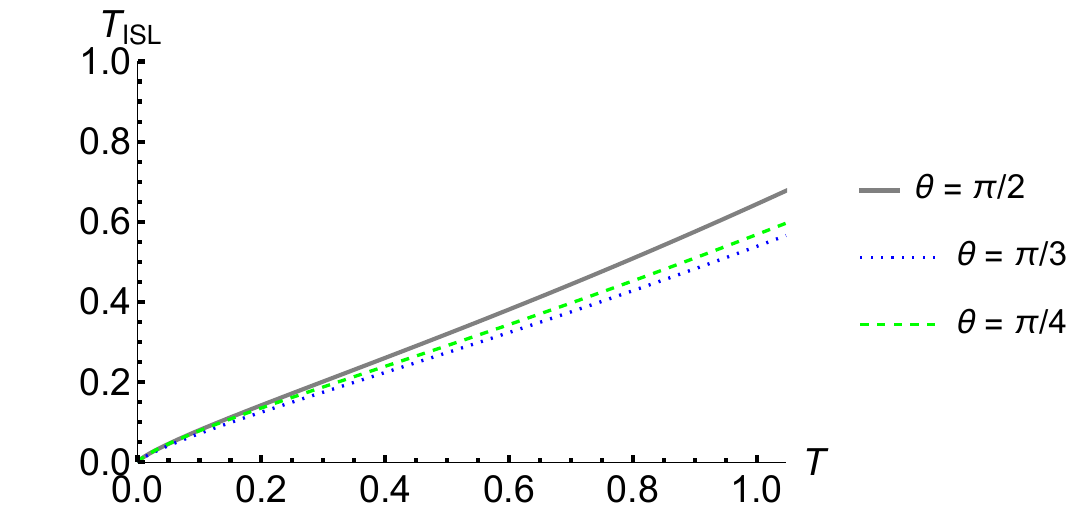}
    \caption{
    The actual evolution time $T$ under the dissipative channel is compared to the speed limit time $T_{\mathrm{ISL}}$  defined in \eqref{geometricisl} and the dynamics comes from the
master equation specified by~\eqref{equa} with $\gamma_t = 2$. The black solid, blue dotted, and green dashed lines represent the initial state parameters $\theta = \frac{\pi}{2}$, $\theta = \frac{\pi}{3}$, and $\theta = \frac{\pi}{4}$, respectively. All times are in units of $(\gamma_t)^{-1}$ (the horizontal axis shows the dimensionless time $\gamma_t t$).
   }
    \label{dissipative _fig4}
\end{figure}

{\textit{Dissipative dynamics}}.-- Consider the Hamiltonian system (\ref {dissipative Ham}) with the master equation (\ref{equa}). With the initial state (\ref{ini_state}), the density matrix $\rho_t$ is given by (\ref{dissipativestate}), see \textbf{Appendix C} for numerical calculations. We study the $T_{\mathrm{ISL}}$ dependence on the evolution time $T$ numerically, see Fig. \ref{dissipative _fig4}.


To facilitate a direct comparison, we juxtapose the ISL based on the relative entropy measure (\ref{isl}) with the ISL based on the geometric measure [Eq.~(\ref{geometricisl})] under dissipative dynamics. As displayed in Figs.~\ref{dissipative _fig2} and \ref{dissipative _fig4}, the geometric bound is generally tighter at a fixed evolution time, although the precise ordering depends on the initial angle and is clarified
below.

\textit{Remark.}
In the dissipative model, the ISLs obtained from the relative entropy measure [Eq.~(\ref{isl}), Fig.~\ref{dissipative _fig2}] and from the geometric measure [Eq.~(\ref{geometricisl}), Fig.~\ref{dissipative _fig4}] may order the speed differently depending on the initial angle $\theta$. For $\theta=\pi/2$ (strong imaginarity), the ISL in Fig.~\ref{dissipative _fig2} exceeds that in Fig.~\ref{dissipative _fig4}, indicating that the entropy--based ISL is tighter. The dominant bottleneck is the information--theoretic and irreversible cost of erasing imaginarity. In contrast, for $\theta=\pi/3,\pi/4$ (weaker imaginarity), the ISL in Fig.~\ref{dissipative _fig4} exceeds that in Fig.~\ref{dissipative _fig2}, showing that the geometric distance to the real subspace is more sensitive in the near--real regime and thus yields a tighter lower bound. This measure--dependent ordering is consistent with resource--theory intuition (entropic cost versus metric distance) and suggests that the choice of measure should be aligned with the dominant dynamical feature of the trajectory.

\textit{Imaginarity speed limit from Liouville space.}  Recent work~\cite{arXiv2406.08584} shows that, in the Liouville space, the instantaneous speed of evolution is governed by the fluctuation (variance) \(\Delta\mathcal L_t\) of the Liouvillian, with \(\Delta\mathcal L_t \le \|\mathcal L_t\|\), where \(\|\cdot\|\) denotes the operator norm
induced by the Hilbert--Schmidt inner product. In particular, when the Liouvillian is time--independent, this yields a time--independent speed bound. Building on this observation, we now prove the following imaginarity speed limit, which relates the change in geometric imaginarity to the time-average of \(\Delta\mathcal L_t\). And we also aim to obtain a time--independent lower bound on the speed limits of imaginarity.

To bound how fast imaginarity can change under a CPTP dynamics, we work in the vectorized representation of density operators and view the generator as a Liouvillian superoperator.  We establish the following Theorem~\ref{thm4} and Corollary~\ref{cor1}, see \textbf{Appendix~F} for detailed derivations.
\begin{theorem}
\label{thm4}
For a CPTP dynamics generated by a (possibly time-dependent) Liouvillian $\mathcal{L}_t$, the evolution time $T$ required to drive the system from the initial state $\rho_0$ to the final state $\rho_T$ is bounded from below by
\begin{align}
\label{f2}
T \;\ge\; T_{\mathrm{ISL}}
       \;=\; \frac{\left|\Delta_I\right|}{\Lambda_{\Delta \mathcal{L}}(T)},
\end{align}
where $\Delta_I= \arccos\sqrt{1-\mathscr{M}_g(\rho_T)}- \arccos\sqrt{1-\mathscr{M}_g(\rho_0)},$
$\Lambda_{\Delta \mathcal{L}}(T)= \frac{1}{T}\int_0^T \! dt \; \Delta\mathcal{L}_t,\qquad
(\Delta \mathcal{L}_t)^2= \mathrm{tr}\!\big(\mathcal{L}_t^{\dagger}\mathcal{L}_t\,\mathcal{P}_t\big)
   - \mathrm{tr}\!\big(\mathcal{L}_t^{\dagger}\mathcal{P}_t\big)\,
     \mathrm{tr}\!\big(\mathcal{L}_t\,\mathcal{P}_t\big),$
and $\mathcal{P}_t=|\tilde{\rho}_t)(\tilde{\rho}_t|$ with $|\tilde{\rho}_t)=|\rho_t)/\sqrt{\mathrm{tr}(\rho_t^2)}$ the normalized Liouville--space state vector associated with $\rho_t$.
\end{theorem}


The quantity $\Delta\mathcal L_t$ can be bounded above by an operator norm, which leads to a bound that does not require solving the dynamics.  In particular, when the generator is time independent we obtain a  time--independent lower bound.

\begin{corollary}\label{cor1}
If $\mathcal L_t\equiv\mathcal L$ is time independent, then $\Delta\mathcal L_t\le \|\mathcal L\|$ for all $t$, where $\|\mathcal L\|:=\sup_{(\psi|\psi)=1}\sqrt{(\psi|\mathcal L^\dagger\mathcal L|\psi)}$
is the operator norm induced by the Hilbert--Schmidt inner product.  Hence,
\[
T \;\ge\; \frac{|\Delta_I|}{\|\mathcal L\|}.
\]
\end{corollary}
We connect QSLs to the resource theory of quantum imaginarity. Our analysis provides both (i) an exact, dynamics--dependent lower bound (Theorem~\ref{thm4}) and (ii) for a time--independent Liouvillian, a time--independent speed limit that relies only on the generator and does not require solving the dynamics (Corollary~\ref{cor1}).

{\textit{QSL in stochastic-approximate transformations.}}--Consider any arbitrary state $\rho$ and define a set of states  $S_{\rho, f}$ such that for all $\rho^{\prime} \in S_{\rho, f}$, the fidelity satisfies $F\left(\rho, \rho^{\prime}\right) \geq f$. The minimal geometric measure of imaginarity within the set $S_{\rho, f}$  is given by \cite{TVKondra2021,Sci. China Phys. Mech. Astron.68220311},
$$
\min _{\rho^{\prime} \in S_{\rho, f}} \mathscr{M}_g\left(\rho^{\prime}\right)=\sin ^2\left(\max \left\{\arcsin \sqrt{\mathscr{M}_g(\rho)}-\arccos  \sqrt{f}, 0\right\}\right).
$$
For a specified target fidelity $f$, this expression provides the minimum possible geometric imaginarity among all states $\rho^{\prime}$ such that the fidelity between $\rho$ and $\rho^{\prime}$ is at least $f$.



We now proceed to estimate the QSL for the transformation between $\rho$ and $\rho'$ within the set $ S_{\rho, f} $. Let us assume that $\rho^\star$ and ${\rho'}^\star$ represent the closest real states to $\rho$ and $\rho'$, respectively, based on the Bures angle $\Theta(\rho,\sigma)=\arccos \sqrt{F(\rho,\sigma)}$ and  (\ref{theta1}). By applying the triangle inequality, we obtain the following result
\begin{align*}
\arccos \left(\sqrt{1-\mathscr{M}_g(\rho)}\right)
& =\Theta\left(\rho, \rho^\star\right) \leq \Theta\left(\rho, {\rho'}^\star\right) \\
& \leq \Theta\left(\rho, \rho' \right)+\Theta\left(\rho',{\rho'}^\star\right)\\
&\leq \arccos(\sqrt{f})+\arccos\left(\sqrt{1-\mathscr{M}_g\left(\rho^{\prime}\right)}\right)
\end{align*}
and
\begin{align*}
\arccos\left(\sqrt{1-\mathscr{M}_g\left(\rho'\right)}\right)
& =\Theta\left(\rho', {\rho'}^\star\right) \leq \Theta\left(\rho', \rho^\star\right) \\
& \leq \Theta\left(\rho, \rho'\right)+\Theta\left(\rho, \rho^\star\right) \\
& \leq \arccos(\sqrt{f})+\arccos\left(\sqrt{1-\mathscr{M}_g(\rho)}\right).
\end{align*}
Therefore,
\begin{align*}
&|\Delta_I|=\left|\arccos\sqrt{1-\mathscr{M}_g(\rho)}-\arccos\sqrt{1-\mathscr{M}_g(\rho')}\right|\\
&\leq|\arccos (\sqrt{f})|.
\end{align*}
Consider the dynamical evolution of the state $\rho_t$, starting from the initial state $\rho$ and evolving to the final state $\rho^{\prime}$ over the time interval $t\in[0,T]$. Following a similar approach as in Theorem \ref{thm3}, we derive the following inequality
\begin{align*}
T\geq \frac{\left\vert\arccos(\sqrt{f})\right\vert}{\Lambda_{g}(T)},
\end{align*}
which gives the QSL of the transformation from $\rho$ to $\rho'$.

\section{Conclusion}\label{sect3}

Understanding how quickly the information can be created or erased is crucial for controlling quantum systems in information processing tasks.
Limitations on the rate at which informational variations \textbf{occur} are critical for designing protocols capable of manipulating quantum states in quantum computers and communication devices. In this study, we have introduced and explored a new concept of speed limits based on the variations in imaginarity, offering general bounds on the evolution time of quantum systems.
We investigate these bounds in several illustrative cases. In particular, these ISL bounds are tight in the sense that the trace distance bound (Theorem \ref{thm2}) is saturated by geodesic evolution, and the geometric  bound (Eq. (\ref{geometricisl})) is achieved for maximally imaginary initial states.
Our findings provide QSLs for the evolution of imaginarity, utilizing various measures such as relative entropy, trace distance, and geometric imaginarity. These QSLs set lower bounds on the rate at which a physical system can either acquire or dissipate a specific amount of imaginarity during its evolution.

Complex numbers play a fundamental role in quantum physics and are essential in characterizing quantum states and their dynamical evolutions. The recently formulated resource theory of imaginarity enables a systematic and quantitative investigation on the imaginary components in quantum systems. Since the imaginary part of a quantum state, analogous to the coherence in resource theories, reflects basis-dependent features, the QSL bounds on imaginarity may provide a lower limit on the time required for a system to transition from quantum behavior to classical characteristics. Recent studies have explored the role of quantum imaginarity in QSLs.
For instance, by using two-time correlation functions formulated as Kirkwood--Dirac quasiprobabilities, in Ref. \cite{PratapsiQST2025} the authors derived QSLs for cases where the initial state and the measured observable do not commute. Via the Schr\"{o}dinger--Robertson relation, the results give rise to lower bounds on the time for the real part of the quasiprobability to become negative and for the imaginary part to exceed a threshold, which are signatures of nonclassical dynamics. This approach is complementary to ours: their bounds concern the emergence of KD nonclassicality for a fixed observable, whereas our ISL bounds constrain the rate of change of the state's imaginarity.
In contrast, our work focuses on imaginarity as a state--based resource defined directly from the density matrix, and  we derive QSL bounds based on three different measures of imaginarity, focusing on the generation and degradation of imaginarity under specific quantum dynamical scenarios including dephasing, dissipation and stochastic--approximate transformations. The imaginarity--based QSLs presented here offer a complementary perspective for exploring fundamental limits in quantum dynamics, potentially providing new tools for quantum information processing and the development of quantum technologies.
Our results may find potential applications in quantum computing, quantum communication, quantum control, quantum thermodynamics and related quantum technologies.

To place our approach within the broader landscape of quantum speed limits, we briefly relate it to geometry--based formulations that do not rely on measures of resources. It is clear that our bounds are resource--dependent, linking the attainable speed to the creation or depletion of imaginarity, while the phase distance and gauge distance QSLs ~\cite{SunPRL127_100404,SunZhengPRL123_180403} provide complementary constraints determined by the path in state space. These bounds together may yield tighter and operationally relevant limits in open-system scenarios.

\section*{ACKNOWLEDGEMENTS}
This work is supported by the National Natural Science Foundation of China (NSFC) under Grant Nos. 12171044; the specific research fund of the Innovation Platform for Academicians of Hainan Province.

\section*{APPENDIX}
\section*{  Proof of Theorem 1}
$\mathit{Proof}$.
From $\mathscr{M}_r(\rho_t)= S(\operatorname{Re}\rho_t)-S(\rho_t)$ and Lemma \ref{lem:co-change}, we get
\begin{equation*}
   \frac{{\rm d} }{{\rm d}t}\mathscr{M}_r(\rho_t)= -{\rm tr}[\mathcal{L}_{t}(\operatorname{Re}\rho_t) \ln \operatorname{Re}\rho_t ]+{\rm tr}[\mathcal{L}_{t}({\rho_t}) \ln \rho_t].
\end{equation*}
Taking the absolute value, we obtain:
\begin{equation}
   \left|\frac{{\rm d}}{{\rm d}t} \mathscr{M}_r(\rho_t)\right|\leq |{\rm tr}[\mathcal{L}_{t}(\operatorname{Re}\rho_t) \ln (\operatorname{Re}\rho_t) ]|+|{\rm tr}[\mathcal{L}_{t}({\rho_t}) \ln \rho_t]|.
\end{equation}
By utilizing the Cauchy--Schwarz inequality,  that $|{\rm tr}[NM]|\leq\sqrt{{\rm tr}[N^{\dagger}N]{\rm tr}[M^{\dagger}M]}$, we can derive the following result
\begin{equation*}
   \left|\frac{{\rm d}}{{\rm d}t} \mathscr{M}_r(\rho_t)\right|\leq \norm{\mathcal{L}_{t}({\operatorname{Re}\rho_t})}_{\rm HS} \norm{\ln (\operatorname{Re}\rho_t)}_{\rm HS} +\norm{\mathcal{L}_{t}({\rho_t})}_{\rm HS} \norm{\ln \rho_t}_{\rm HS}.
\end{equation*}
Therefore, integrating over time from 0 to \(T\),
\begin{align*}
  \int_{0}^{T} {\rm d}t\left|\frac{{\rm d} }{{\rm d}t}\mathscr{M}_r(\rho_t)\right|
  &\leq  \int_{0}^{T}  \norm{\mathcal{L}_{t}({\operatorname{Re}\rho_t})}_{\rm HS} \norm{\ln (\operatorname{Re}\rho_t)}_{\rm HS}   {\rm d}t\nonumber \\
  &+\int_{0}^{T}  \norm{\mathcal{L}_{t}({\rho_t})}_{\rm HS} \norm{{\ln \rho_t}}_{\rm HS} {\rm d}t.
\end{align*}
By applying the Cauchy--Schwarz inequality
$\int_{0}^{T}\eta(t) \gamma(t) {\rm d}t \leq \sqrt{\int_{0}^{T} \eta(t)^{2}{\rm d}t} \sqrt{\int_{0}^{T} \gamma(t)^{2}{\rm d}t}$,
we obtain the following result
\begin{align*}
  \int_{0}^{T}{\rm d}t\left|\frac{{\rm d}}{{\rm d}t} \mathscr{M}_r(\rho_t)\right| \leq& \sqrt{\int_{0}^{T}  \norm{\mathcal{L}_{t}({\operatorname{Re}\rho_t)}}^{2}_{\rm HS}{\rm d}t} \sqrt{\int_{0}^{T}  \norm{\ln (\operatorname{Re}\rho_t)}^{2}_{\rm HS} {\rm d}t}\nonumber\\
  &+\sqrt{\int_{0}^{T}  \norm{\mathcal{L}_{t}({\rho_t})}^{2}_{\rm HS}{\rm d}t} \sqrt{\int_{0}^{T}  \norm{{\ln \rho_t}}^{2}_{\rm HS} {\rm d}t}.
\end{align*}
Hence, dividing through by \(T\), we obtain,
\begin{align*}
  &\frac{1}{T}\int_{0}^{T}{\rm d}t\left|\frac{{\rm d}}{{\rm d}t} \mathscr{M}_r(\rho_t)\right|\leq\nonumber\\
  & \sqrt{  \frac{1}{T}\int_{0}^{T}  \norm{\mathcal{L}_{t}({\operatorname{Re}\rho_t)}}^{2}_{\rm HS}{\rm d}t} \sqrt{  \frac{1}{T}\int_{0}^{T}  \norm{\ln (\operatorname{Re}\rho_t)}^{2}_{\rm HS} {\rm d}t}\nonumber\\
  &+\sqrt{  \frac{1}{T}\int_{0}^{T}  \norm{\mathcal{L}_{t}({\rho_t})}^{2}_{\rm HS}{\rm d}t} \sqrt{  \frac{1}{T}\int_{0}^{T}  \norm{{\ln \rho_t}}^{2}_{\rm HS} {\rm d}t}.
\end{align*}
Recognizing that \(\int_0^T \left| \frac{\mathrm{d}}{\mathrm{d}t} \mathscr{M}_r(\rho_t) \right| \mathrm{d}t \geq |\mathscr{M}_r(\rho_T) - \mathscr{M}_r(\rho_0)|\) and substituting the definitions of \(\Lambda_T^{\mathrm{Re}}\) and \(\Lambda_T\), which gives rise to (\ref{isl}). $\Box$

{\section*{Appendix A: The derivation of the state distance $\Theta$}
Set
\begin{equation*}
  f(\rho_t;k)=\sqrt{F(\rho_t,\rho_{k})}=\operatorname{tr} \sqrt{\sqrt{\rho_t} \rho_{k} \sqrt{\rho_t}}.
\end{equation*}
When $k$ is close to $t$, we expand $f(\rho_t;k)$ to the second order with respect to $k$, \begin{equation}
  f(\rho_t;k)=f(\rho_t,t)+\frac{\partial}{\partial k}f(\rho_t;k)\Big\vert_{k=t}dt+\frac{1}{2}\frac{\partial^2}{\partial k^2}f(\rho_t;k)\Big\vert_{k=t}dt^2
\end{equation}
with $dt=k-t$. It is verified that
$$ f(\rho_t;t)=\operatorname{tr}\sqrt{\sqrt{\rho_t} \rho_{t} \sqrt{\rho_t}}=\operatorname{tr}\rho_t=1,$$
\begin{align*}
\frac{\partial}{\partial k}f(\rho_t;k)\Big\vert_{k=t}=\frac{\partial}{\partial k}\operatorname{tr}\sqrt{\sqrt{\rho_t} \rho_{k} \sqrt{\rho_t}}\Big\vert_{k=t}=\operatorname{tr}\sqrt[4]{\rho_t}\left(\frac{d}{dt}{\sqrt{\rho_t}}\right)\sqrt[4]{\rho_t}=0
\end{align*}
and
\begin{align*}
&\frac{\partial^2}{\partial k^2}f(\rho_t;k)\Big\vert_{k=t}=\frac{\partial^2}{\partial k^2}\operatorname{tr}\sqrt{\sqrt{\rho_t} \rho_{k} \sqrt{\rho_t}}\Big\vert_{k=t}\\
&=\operatorname{tr}\sqrt[4]{\rho_t}\left(\frac{d^2}{dt^2}\sqrt{\rho_t}\right)\sqrt[4]{\rho_t}
=-\operatorname{tr}\left(\frac{d}{dt}{\sqrt{\rho_t}}\right)^2. \end{align*}
Hence,
\begin{align*}
  f(\rho_t;t+dt)=\operatorname{tr} \sqrt{\sqrt{\rho_t} \rho_{k} \sqrt{\rho_t}}=1-\frac{1}{2}\operatorname{tr}\left(\frac{d}{dt}{\sqrt{\rho_t}}\right)^2
\end{align*}
when $k$ approaches $t$. Therefore,
\begin{align}
f(\rho_t;t+dt)&=1-\frac{1}{2}\operatorname{tr}\left(\frac{d}{dt}{\sqrt{\rho_t}}\right)^2\label{thetaequality1}\\
&=\cos\Theta(\rho_t,\rho_{t+dt})\label{thetaequality2}\\
&=1-\frac{1}{2}\Theta^2(\rho_t,\rho_{t+dt}),\label{thetaequality3}
\end{align}
here the expression in (\ref{thetaequality3}) follows from the Taylor expansion of the $\cos$  function, which is given by \( \cos x = 1 - \frac{x^2}{2} + \frac{x^4}{24} + o(x^4) \).

From (\ref{thetaequality1}) and (\ref{thetaequality3}), we obtain
\begin{align}
  \Theta^2(\rho_t,\rho_{t+dt})=\operatorname{tr}\left(\frac{d}{dt}{\sqrt{\rho_t}}\right)^2.
\end{align}

{\section*{Appendix B: Derivation of the ISL bound (\ref{geometricisl}) for geometric imaginarity in dephasing dynamics}

\textbf{Case 1, $\theta= \frac{\pi}{2}$}. The initial state is
\begin{equation}
\rho_0=\left(\begin{array}{ll}
\frac{1}{2} & \frac{-i }{2}\\
\frac{i}{2}& \frac{1}{2}
\end{array}\right)\label{B1}
\end{equation}
and
\begin{equation}
    \rho_t=\left(\begin{array}{ll}
       \frac{1}{2} & -i\frac{1}{2}e^{-2t}\\
      i\frac{1}{2}e^{-2t} & \frac{1}{2}
    \end{array}\right).\label{B2}
\end{equation}
$\sqrt{\rho_t}$ is given by
\begin{equation}
\left(\begin{array}{cc}
\frac{\sqrt{e^{-r t}\left(-1+e^{2 t}\right)}}{2 \sqrt{2}}+\frac{\sqrt{e^{-2 t}\left(1+e^{2 t}\right)}}{2 \sqrt{2}} & \frac{i \sqrt{e^{-2 t}\left(-1+e^{2 t}\right)}}{2 \sqrt{2}}-\frac{i \sqrt{e^{-2 t}\left(1+e^{2 t}\right)}}{2 \sqrt{2}} \\
-\frac{i \sqrt{e^{-2 t}\left(-1+e^{2 t}\right)}}{2 \sqrt{2}}+\frac{i \sqrt{e^{-2 t}\left(1+e^{2 t}\right)}}{2 \sqrt{2}} & \frac{\sqrt{e^{-2 t}\left(-1+e^{2 t}\right)}}{2 \sqrt{2}}+\frac{\sqrt{e^{-2 t}\left(1+e^{2 t}\right)}}{2 \sqrt{2}}
\end{array}\right).\label{B4}
\end{equation}
Then
\begin{equation}
\frac{d}{dt}{\sqrt{\rho_t}}=
\left(\begin{array}{cl}\frac{e^{-2 t}\left(\frac{1}{\sqrt{1-e^{-2 t}}}-\frac{1}{\sqrt{1+e^{-2 t}}}\right)}{2 \sqrt{2}} & \frac{i e^{-2 t}\left(\frac{1}{\sqrt{1-e^{-2 t}}}+\frac{1}{\sqrt{1+e^{-2 t}}}\right)}{2 \sqrt{2}} \\ -\frac{i e^{-2 t}\left(\frac{1}{\sqrt{1-e^{-2 t}}}+\frac{1}{\sqrt{1+e^{-2 t}}}\right)}{2 \sqrt{2}} & \frac{e^{-2 t}\left(\frac{1}{\sqrt{1-e^{-2 t}}}-\frac{1}{\sqrt{1+e^{-2 t}}}\right)}{2 \sqrt{2}}\end{array}\right)\label{B5}
\end{equation}
and $(\frac{d}{dt}{\sqrt{\rho_t}})^2$ is of the form
\small
\begin{equation}
\frac{1}{4\sqrt{2}}
{\left(
\begin{array}{cc}
 a+b & -\frac{i e^{-6 t} \sqrt{-e^{8 t} \left(2 e^{4 t} \left(\sqrt{1-e^{-4 t}}-1\right)+1\right)} (a-b)}{\sqrt{1-e^{-4 t}}-1} \\
 \frac{i e^{6 t} \left(\sqrt{1-e^{-4 t}}-1\right) (a-b)}{\sqrt{-e^{8 t} \left(2 e^{4 t} \left(\sqrt{1-e^{-4 t}}-1\right)+1\right)}} & a+b \\
\end{array}
\right)}\label{B6},
\end{equation}
where $a=\sqrt{\sqrt{1-e^{-4 t}}-e^{-8 t} \sqrt{e^{8 t} \left(-2 e^{4 t} \left(\sqrt{1-e^{-4 t}}-1\right)-1\right)}+1}$ and $b=\sqrt{\sqrt{1-e^{-4 t}}+e^{-8 t} \sqrt{e^{8 t} \left(-2 e^{4 t} \left(\sqrt{1-e^{-4 t}}-1\right)-1\right)}+1}$.

Moreover,
\begin{align*}
&|\Delta_I|= \left|\arccos\sqrt{1-\mathscr{M}_g(\rho_T)}-\arccos\sqrt{1-\mathscr{M}_g(\rho_0)}\right|\\
&= \left|\arccos\frac{1+\sqrt{F\left(\rho_T, {\rho_T}^T\right)}}{2}-\arccos\frac{1+\sqrt{F\left(\rho_0, {\rho_0}^T\right)}}{2}\right|\\
&= \left|\arccos\frac{1+\operatorname{tr} \sqrt{\sqrt{\rho_T} {\rho_T}^T \sqrt{\rho_T}}}{2}-\arccos\frac{1+\operatorname{tr} \sqrt{\sqrt{\rho_0} {\rho_0}^T \sqrt{\rho_0}}}{2}\right|,
\end{align*}
where the second equality is based on the definition of the geometric imaginarity (\ref{geometric1}), the thrid equality is due to that $\sqrt{F(\rho_T, {\rho_T}^T)}=\operatorname{tr} \sqrt{\sqrt{\rho_T} {\rho_T}^T \sqrt{\rho_T}}$ and $\sqrt{F(\rho_0, {\rho_0}^T )}=\operatorname{tr} \sqrt{\sqrt{\rho_0} {\rho_0}^T \sqrt{\rho_0}}$.
By using the equations in(\ref{B1})-(\ref{B5}), we get
\begin{align*}
&|\Delta_I|&= \left| \cos ^{-1}\left(\frac{3}{4}\right)-\cos ^{-1}\left(\frac{1}{8} \left(\sqrt{2} \left(\sqrt{\alpha-\beta}+\sqrt{\alpha+\beta}\right)+4\right)\right) \right|,
\end{align*}
where $\alpha= \sqrt{1-e^{-4 T}}+1$ and $\beta=e^{-8 T} \sqrt{e^{8 T} \left(-2 e^{4 T} \left(\sqrt{1-e^{-4 T}}-1\right)-1\right)}$.

From (\ref{B6}), we have
\begin{align*}
{\Lambda_{g}(T)}=\frac{1}{T}\int_0^T dt
\sqrt{\mathrm{tr}\left(\frac{d}{dt}\sqrt{\rho_t}\right)^2}
=\frac{1}{T}{\int_0^T \sqrt{\frac{1}{e^{4 t}-1}} \, dt},
\end{align*}
where $\mathrm{tr}(\frac{d}{dt}{\sqrt{\rho_t}})^2= \frac{1}{e^{4 t}-1}$. Thus,
\begin{align*}
&T\geq T_{\mathrm{ISL}}=\frac{\left\vert\Delta_I\right\vert}{\Lambda_{g}(T)}\\
&=\frac{\left| \cos ^{-1}\left(\frac{3}{4}\right)-\cos ^{-1}\left(\frac{1}{8} \left(\sqrt{2} \left(\sqrt{\alpha-\beta}+\sqrt{\alpha+\beta}\right)+4\right)\right) \right|}{\frac{1}{T}{\int_0^T \sqrt{\frac{1}{e^{4 t}-1}} \, dt}}.
\end{align*}

\textbf{Case 2, $\theta= \frac{\pi}{3}$}. The initial state is
\begin{equation}
\rho_0=\left(
\begin{array}{cc}
 \frac{3}{4} & -\frac{i\sqrt{3}}{4}  \\
 \frac{i \sqrt{3}}{4} & \frac{1}{4} \\
\end{array}
\right)\label{B11}
\end{equation}
and the state $\rho_t$ is given by
\begin{equation}
    \rho_t=\left(
\begin{array}{cc}
 \frac{3}{4} & -\frac{ \sqrt{3}}{4} i e^{-2 t} \\
 \frac{ \sqrt{3}}{4} i e^{-2 t} & \frac{1}{4} \\
\end{array}
\right).\label{B12}
\end{equation}
Then $\sqrt{\rho_t}$ is given by
\begin{equation}\left(
\begin{array}{cc}
 \frac{\left(a+e^{4 t}\right) \sqrt{a e^{-4 t}+2}+\left(a-e^{4 t}\right) \sqrt{2-a e^{-4 t}}}{4 a} & -\frac{i e^{2 t} \left(\sqrt{2-a e^{-4 t}}-\sqrt{a e^{-4 t}+2}\right)}{4 \sqrt{e^{4 t}+\frac{e^{8 t}}{3}}} \\
 \frac{i e^{2 t} \left(\sqrt{2-a e^{-4 t}}-\sqrt{a e^{-4 t}+2}\right)}{4 \sqrt{e^{4 t}+\frac{e^{8 t}}{3}}} & \frac{\left(a-e^{4 t}\right) \sqrt{a e^{-4 t}+2}+\left(a+e^{4 t}\right) \sqrt{2-a e^{-4 t}}}{4 a} \\
\end{array}
\right)
,\label{B14}
\end{equation}
where $a=\sqrt{e^{4 t} \left(e^{4 t}+3\right)}$.
And $\frac{d}{dt}{\sqrt{\rho_t}}$ has the form
\small{\begin{equation}
\left(
\begin{array}{cc}
 \sqrt{3} e^{4 t} \left(3 a e^{4 t}+\text{ak}-3 u\right) & -i e^{2 t} \left(2 a k e^{4 t}+3 a k+4 e^{8 t} u\right) \\
 i e^{2 t} \left(3 a k+2 k \text{ae}^{4 t}+4 e^{8 t} u\right) & \sqrt{3} e^{4 t} \left(-a k-5 e^{4 t} u-3 u\right) \\
\end{array}
\right),\label{B15}
\end{equation}}
where $b=e^{-4 t} \sqrt{e^{4 t} \left(e^{4 t}+3\right)}$, $c=4 \sqrt{1-e^{-4 t}} \left(e^{4 t} \left(e^{4 t}+3\right)\right)^{3/2}$, $a=\sqrt{2-b}+\sqrt{b+2}$, $u=\sqrt{2-b}-\sqrt{b+2}$ and $k=\sqrt{e^{4 t} \left(e^{4 t}+3\right)}$.

From (\ref{B11})-(\ref{B15}), we get
\small{\begin{align*}
&|\Delta_I|= \\
&\left|  \cos ^{-1}\left(\frac{\sqrt{7}}{8}+\frac{1}{2}\right)-\cos ^{-1}\left(\frac{\sqrt{-\alpha+\beta-\gamma+10}+\sqrt{\alpha+\beta-\gamma+10}}{8 \sqrt{2}}+\frac{1}{2}\right)\right|,
\end{align*}}
where $\alpha= \sqrt{\frac{e^{-8 T} \left(e^{4 T} \left(-36 \sqrt{3-3 e^{-4 T}}+64 e^{8 T}+48 e^{4 T} \left(\sqrt{3-3 e^{-4 T}}+2\right)+63\right)-27\right)}{e^{4 T}+3}}$, $\beta=\frac{6 \sqrt{3-3 e^{-4 T}}}{e^{4 T}+3}$ and $\gamma=\frac{12}{e^{4 T}+3}$.

Since
\begin{align*}
{\Lambda_{g}(T)}=\frac{1}{T}\int_0^T dt
\sqrt{\mathrm{tr}\left(\frac{d}{dt}\sqrt{\rho_t}\right)^2}
=\frac{1}{T}{\int_0^T  \sqrt{\frac{2}{2 e^{4 t}-2}} \, dt},
\end{align*}
where $\mathrm{tr}(\frac{d}{dt}{\sqrt{\rho_t}})^2= \frac{2}{2 e^{4 t}-2}$,
we obtain
\begin{align*}
&T\geq T_{\mathrm{ISL}}=\frac{\left\vert\Delta_I\right\vert}{\Lambda_{g}(T)}\\
&=\frac{\left|  \cos ^{-1}\left(\frac{\sqrt{7}}{8}+\frac{1}{2}\right)-\cos ^{-1}\left(\frac{\sqrt{-\alpha+\beta-\gamma+10}+\sqrt{\alpha+\beta-\gamma+10}}{8 \sqrt{2}}+\frac{1}{2}\right)\right|}{\frac{1}{T}{\int_0^T  \sqrt{\frac{2}{2 e^{4 t}-2}}\, dt}}.
\end{align*}

\textbf{Case 3, $\theta= \frac{\pi}{4}$}. The initial state is
\begin{equation}
\rho_0=\left(
\begin{array}{cc}
 \frac{\sqrt{2}+2}{4}  & -\frac{i }{2 \sqrt{2}} \\
 \frac{i }{2 \sqrt{2}} & \frac{ 2-\sqrt{2}}{4} \\
\end{array}
\right),\label{B31}
\end{equation}
and the state $\rho_t$ is given by
\begin{equation}
    \rho_t=\left(
\begin{array}{cc}
 \frac{\left(\sqrt{2}+2\right)}{4}  & -\frac{i e^{-2 t}}{2 \sqrt{2}} \\
 \frac{i e^{-2 t}}{2 \sqrt{2}} & \frac{ \left(2-\sqrt{2}\right)}{4} \\
\end{array}
\right).\label{B32}
\end{equation}
Similar to the calculations for \textbf{Cases 1 and 2}, we have
\small{\begin{align*}
&|\Delta_I|=\\
& \left|\cos ^{-1}\left(\frac{\sqrt{\frac{5}{2}}}{4}+\frac{1}{2}\right)-\cos ^{-1}\left(\frac{1}{8} \left(\sqrt{\alpha-\beta}+\sqrt{\alpha+\beta}\right)+\frac{1}{2}\right) \right|,
\end{align*}}
where $\alpha= \frac{\sqrt{2-2 e^{-4 T}}+6 e^{4 T}+4}{e^{4 T}+1}$ and 

$\beta=\frac{e^{-8 T} \sqrt{e^{8 T} \left(e^{4 T}+1\right)^3 \left(e^{4 T} \left(-2 \sqrt{2-2 e^{-4 T}}+8 e^{4 T} \left(\sqrt{2-2 e^{-4 T}}+4 e^{4 T}+2\right)+3\right)-1\right)}}{\left(e^{4 T}+1\right)^2}$,
and
\small{\begin{align*}
&{\Lambda_{g}(T)}=\\
&\frac{1}{T}{\int_0^T \sqrt{\frac{e^{6 t} \left(\left(3-2 \sqrt{2-2 e^{-4 t}}\right) \sinh (2 t)+\cosh (2 t)\right)+1}{\left(e^{4 t}-1\right) \left(e^{4 t}+1\right)^2}}  \, dt}.
\end{align*}}
Finally, we obtain
\small{\begin{align*}
&T\geq T_{\mathrm{ISL}}=\frac{\left\vert\Delta_I\right\vert}{\Lambda_{g}(T)}\\
&=\frac{\left|\cos ^{-1}\left(\frac{\sqrt{\frac{5}{2}}}{4}+\frac{1}{2}\right)-\cos ^{-1}\left(\frac{1}{8} \left(\sqrt{\alpha-\beta}+\sqrt{\alpha+\beta}\right)+\frac{1}{2}\right) \right|}{\frac{1}{T}{\int_0^T \sqrt{\frac{e^{6 t} \left(\left(3-2 \sqrt{2-2 e^{-4 t}}\right) \sinh (2 t)+\cosh (2 t)\right)+1}{\left(e^{4 t}-1\right) \left(e^{4 t}+1\right)^2}}\, dt}}.
\end{align*}}

{\section*{Appendix C: The ISL bound (\ref{geometricisl}) for geometric imaginarity in dissipative dynamics}

Using similar calculations to the ones in \textbf{Appendix B}, we obtain the following:

\textbf{Case 1, $\theta= \frac{\pi}{2}$}.
\small{
\begin{align*}
T\geq T_{\mathrm{ISL}}
=\frac{\left|\cos ^{-1}\left(\frac{3}{4}\right)-\cos ^{-1}\left(\frac{\sqrt{\beta-\gamma}+\sqrt{\beta+\gamma}}{4 \sqrt{2}}+\frac{1}{2}\right) \right|}{\frac{1}{T}{\int_0^T \frac{1}{4} \sqrt{\frac{e^{\frac{5 t}{2}} \left(\sinh \left(\frac{t}{2}\right) (-8 \alpha \cosh (t)-8 \alpha+26 \cosh (t)+3)+4 \cosh \left(\frac{t}{2}\right)-7 \cosh \left(\frac{3 t}{2}\right)\right)+4}{\left(e^t-1\right) \left(e^t \left(e^t-1\right)+1\right)^2}}\, dt}},
\end{align*}
where $\alpha=\sqrt{e^{-2 t} \left(e^t-1\right)}$, $\beta=4 \sinh (T)-2 \sinh (2 T)-4 \cosh (T)+2 \cosh (2 T)+\frac{\sqrt{e^{-2 T} \left(e^T-1\right)}}{2 \cosh (T)-1}+\frac{1}{1-2 \cosh (T)}+4$ and
$\gamma=\sqrt{\frac{e^{-T} \left(-18 \sqrt{e^{-2 T} \left(e^T-1\right)}-\sinh (T)+\sinh (2 T)+31 \cosh (2 T)+\left(16 \sqrt{e^{-2 T} \left(e^T-1\right)}-111\right) \cosh (T)+81\right)}{2 \cosh (T)-1}}.$
}

\textbf{Case 2, $\theta= \frac{\pi}{3}$}.
\small{
\begin{align*}
T\geq T_{\mathrm{ISL}}
=\frac{\left|\cos ^{-1}\left(\frac{\sqrt{7}}{8}+\frac{1}{2}\right)-\cos ^{-1}\left(\frac{\sqrt{\beta+\gamma-\eta}+\sqrt{\beta+\gamma+\eta}}{8 \sqrt{2}}+\frac{1}{2}\right) \right|}{\frac{1}{T}{\int_0^T \frac{\sqrt{324-3 e^t \left(e^t \left(e^t \left(4 (3 \alpha-5) e^t+24 \alpha+59\right)-9 \alpha-177\right)-27 \alpha+234\right)}}{16 \left(e^t-1\right) \left(e^t \left(4 e^t-9\right)+9\right)^2}\, dt}},
\end{align*}
where $\alpha=\sqrt{e^{-2 t} \left(e^t-1\right)}$,
$\beta=\frac{12}{5 \sinh (T)-13 \cosh (T)+9}$,
$\gamma=6 e^{-2 T} \left(\frac{3 e^{3 T} \sqrt{e^{-2 T} \left(e^T-1\right)}}{e^T \left(4 e^T-9\right)+9}-4 e^T+3\right)+16$ and
$\eta=\sqrt{\frac{4 \left(64 e^T \left(4 e^T-9\right)-135 \sqrt{e^{-2 T} \left(e^T-1\right)}+666\right)}{e^T \left(4 e^T-9\right)+9}+567 e^{-2 T}+144 e^{-T} \left(\sqrt{e^{-2 T} \left(e^T-1\right)}-6\right)}$.
}

\textbf{Case 3, $\theta= \frac{\pi}{4}$}.
\small{
\begin{align*}
T\geq T_{\mathrm{ISL}}
=\frac{\left|\cos ^{-1}\left(\frac{\sqrt{\frac{5}{2}}}{4}+\frac{1}{2}\right)-\cos ^{-1}\left(\frac{1}{8} \left(\sqrt{\beta-\gamma }+\sqrt{\beta+\gamma }\right)+\frac{1}{2}\right) \right|}{\frac{1}{4T}{\int_0^T \sqrt{\frac{e^t \left({K_1}+e^t \left({K_2}-{K_3} e^t\right)\right)+48 \sqrt{2}+68}{\left(e^t-1\right) \left(e^t \left(2 e^t-2 \sqrt{2}-3\right)+2 \sqrt{2}+3\right)^2}}\, dt}},
\end{align*}
where $\alpha = \sqrt{e^(-2 t) (-1 + e^t)} $, ${K_ 1} = (7\sqrt {2} + 10) \alpha- 100\sqrt {2} - 142$, ${K_ 2} = (\sqrt {2} + 2) \alpha+76\sqrt {2} + 107,  {K_ 3} =2 ((\sqrt {2} + 2) \alpha - 2\sqrt {2} - 5) e^t + (6\sqrt {2} + 8) \alpha + 24\sqrt {2} + 37 $,
$\beta=2 e^{-2 T} \left(e^T \left(-\frac{1}{\sqrt{2} \sqrt{e^{-2 T} \left(e^T-1\right)}+2 \sqrt{e^{-2 T} \left(e^T-1\right)}+2}-2 \sqrt{2}-4\right)+2 \sqrt{2}+3\right)+8$,
$\gamma = \sqrt{\frac{e^{-2 T} \left(A_3+382 \sqrt{2}+541\right)}{e^T \left(2 e^T-2 \sqrt{2}-3\right)+2 \sqrt{2}+3}}$, and
$A_1=\sqrt{2} \sqrt{e^{-2 T} \left(e^T-1\right)}+8 e^T+2 \sqrt{e^{-2 T} \left(e^T-1\right)}-16 \sqrt{2}-30$ $A_2=2 e^T \left(8 A_1 e^T-33 \sqrt{2} \sqrt{e^{-2 T} \left(e^T-1\right)}-50 \sqrt{e^{-2 T} \left(e^T-1\right)}+368 \sqrt{2}+545\right)$, $A_3=e^T \left(80 \sqrt{e^{-2 T} \left(e^T-1\right)}+56 \sqrt{2} \sqrt{e^{-2 T} \left(e^T-1\right)}-1229-862 \sqrt{2}+A_2\right)$.
}
\section*{Appendix D: Closed-form expressions for dephasing dynamics used in the evaluation of Eq.~(\ref{isl})}
By calculation we have
\small{
\begin{align}
&|\mathscr{M}_r(\rho_{t})-\mathscr{M}_r(\rho_{0})|=-\nonumber\\
&\frac{1}{\ln4}\{ \ln \left(1-e^{-\int_0^t dt'\gamma_{t'}-i\omega_0t} \sqrt{e^{2 i\omega_0t} \left(e^{2\int_0^t dt'\gamma_{t'}} \cos ^2\theta +2 (\cos \theta-1) i\cos^2\frac{\theta }{2}\right)}\right)\nonumber\\
&- \ln \left(1+e^{-\int_0^t dt'\gamma_{t'}-i\omega_0t} \sqrt{e^{2 i\omega_0t} \left(e^{2\int_0^t dt'\gamma_{t'}} \cos ^2\theta +2 (\cos \theta-1) i\cos^2\frac{\theta }{2}\right)}\right)\nonumber\\
&+2 \sqrt{e^{2 i\omega_0t}} e^{-\int_0^t dt'\gamma_{t'}-i\omega_0t} \sqrt{e^{2\int_0^t dt'\gamma_{t'}} \cos^2\theta+2 (\cos\theta-1) i\cos^2\frac{\theta }{2}}\nonumber\\
& \coth ^{-1}\left(\frac{e^{\int_0^t dt'\gamma_{t'}+i\omega_0t}}{\sqrt{e^{2 i\omega_0t}} \sqrt{e^{2 \int_0^t dt'\gamma_{t'}} \cos ^2\theta +2 (\cos\theta-1) i\cos\left(\frac{\theta }{2}\right)^2}}\right)-\ln4\},
\label{eq:D1}
\end{align}
}
\begin{align}
\norm{\mathcal{L}(\rho_{t})}_{HS}^{2}=\frac{1}{2} |\gamma^2_t -\omega^2_0|e^{-2\int_0^t dt'\gamma_{t'}+i\omega_0t}\sin ^2\theta,\label{eq:D2}
\end{align}
\begin{align}
\norm{\mathcal{L}(\operatorname{Re}\rho_{t})}_{HS}^{2}=0,\label{eq:D3}
\end{align}
\begin{align}
\norm{\ln{\operatorname{Re}\rho_{t}}}_{HS}^{2}=\left[\ln \left(\sin ^2\frac{\theta }{2}\right)\right]^2+\left[\ln \left(\cos ^2\frac{\theta }{2}\right)\right]^2,\label{eq:D4}
\end{align}
\small{
\begin{align}
&\norm{\ln{\rho_{t}}}_{HS}^{2}=\nonumber\\
&\left[\ln \frac{1-e^{-\int_0^t dt'\gamma_{t'}-i\omega_0t} \sqrt{e^{2 i\omega_0t} \left(e^{2\int_0^t dt'\gamma_{t'}} \cos ^2\theta +2 (\cos \theta-1) i\cos^2\frac{\theta }{2}\right)}}{2} \right]^2\nonumber\\
&+ \left[\ln \frac{1+e^{-\int_0^t dt'\gamma_{t'}-i\omega_0t} \sqrt{e^{2 i\omega_0t} \left(e^{2\int_0^t dt'\gamma_{t'}} \cos ^2\theta +2 (\cos \theta-1) i\cos^2\frac{\theta }{2}\right)}}{2} \right]^2.\label{eq:D5}
\end{align}
}

\section*{Appendix E: Closed-form expressions for dissipative dynamics used in the evaluation of Eq.~(\ref{isl})}
By calculation we have
\small{
\begin{align}
\mathscr{M}_r(\rho_{0})=-\sin ^2\frac{\theta }{2} \ln \left(\sin ^2\frac{\theta }{2}\right)-\cos ^2\frac{\theta }{2} \ln  \left(\cos ^2\frac{\theta }{2}\right),\label{eq:E1}
\end{align}
\begin{align}
& \mathscr{M}_r(\rho_{t})=- e^{-\int_0^t dt'{\frac{\gamma_t'}{2}}}\cos ^2\frac{\theta }{2} \ln \left( e^{-\int_0^t dt'{\frac{\gamma_t'}{2}}}\cos ^2\frac{\theta }{2}\right)\nonumber\\
&-\left(1-e^{-\int_0^t dt'{\frac{\gamma_t'}{2}}}\cos ^2\frac{\theta }{2} \right) \ln \left(1-e^{-\int_0^t dt'{\frac{\gamma_t'}{2}}}\cos ^2\frac{\theta }{2} \right)\nonumber\\
&+\frac{1}{2} \left(\ln \left(\frac{1}{16} \left(\sqrt{2\alpha} e^{-\int_0^t dt'{\frac{\gamma_t'}{2}} }+2\right)\right)+\ln \left(2-\sqrt{2\alpha}  e^{-\int_0^t dt'{\frac{\gamma_t'}{2}}}\right)\right)\nonumber\\
 &+ \sqrt{\frac{{\alpha}}{{2}}}  e^{-\int_0^t dt'{\frac{\gamma_t'}{2}} } \tanh ^{-1}\left(e^{-\int_0^t dt'{\frac{\gamma_t'}{2}} } \sqrt{e^{-\int_0^t dt'{\gamma_t'}}-4\left(e^{\int_0^t dt'{\frac{\gamma_t'}{2}}}-1\right) \cos ^4\frac{\theta }{2} }\right),\label{eq:E2}
\end{align}
\begin{align}
\left\|\mathcal{L}\left(\rho_t\right)\right\|_{H S}^2=\frac{1}{32} {\gamma_t}^2 e^{-\int_0^t dt'{\gamma_t'}}\left(16 \cos ^4 \frac{\theta}{2}+e^{\int_0^t dt'{\frac{\gamma_t'}{2}}} \sin ^2 \theta\right),\label{eq:E3}
\end{align}
\begin{align}
 \left\|\mathcal{L}\left(\operatorname{Re}\rho_t\right)\right\|_{H S}^2=\frac{1}{2} {\gamma_t}^2 e^{-\int_0^t dt'{\gamma_t'}} \cos ^4 \frac{\theta}{2},\label{eq:E4}
\end{align}
\begin{align}
&\norm{\ln{\operatorname{Re}\rho_t}}_{HS}^{2}=\left[\ln\left(e^{-\int_0^t dt'{\frac{\gamma_t'}{2}}}\cos ^2\frac{\theta }{2} \right)\right]^2\nonumber\\
&+\left[\ln\left(1- e^{-\int_0^t dt'{\frac{\gamma_t'}{2}}}\cos ^2\frac{\theta}{2}\right)\right]^2,\label{eq:E5}
\end{align}
\begin{align}
&\norm{\ln{\rho_{t}}}_{HS}^{2}=\left[\ln\left(\frac{1}{4} \left(2-\sqrt{2} e^{-\int_0^t dt'{\frac{\gamma_t'}{2} }}\sqrt{\beta}\right)\right)\right]^2\nonumber\\
&\qquad+ \left[\ln\left(\frac{1}{4} \left(2+\sqrt{2} e^{-\int_0^t dt'{\frac{\gamma_t'}{2} }} \sqrt{\beta}\right)\right)\right]^2,\label{eq:E6}
\end{align}
where $\alpha=3+4 \cos \theta +\cos 2 \theta -8 e^{\int_0^t dt'{\frac{\gamma_t'}{2} }}\cos ^4\frac{\theta }{2}+2 e^{\int_0^t dt'{\gamma_t'}}$ and $\beta=3-4 \left(e^{\int_0^t dt'{\frac{\gamma_t'}{2} }}-1\right)\cos\theta  -\left(e^{\int_0^t dt'{\frac{\gamma_t'}{2} }}-1\right)\cos 2\theta -3 e^{\int_0^t dt'{\frac{\gamma_t'}{2} }}+2 e^{\int_0^t dt'{\gamma_t'}}$.
}

\section*{Appendix F:Liouville-Space Formalism and Detailed Proofs}
Before presenting the proof of Theorem~\ref{thm4}, we first review the following preliminaries.

For every linear operator \(X\) acting on a \(d\)--dimensional Hilbert space \(\mathcal H_d\), there exists a vector
\[
X=\sum_{i j}\alpha_{ij}\,|i\rangle\!\langle j|
\;\longrightarrow\;
|X)=\sum_{i j}\alpha_{ij}\,|j)\otimes|i)
=\sum_{i j}\alpha_{ij}\,|ji).
\]
These vectorized linear operators form a Hilbert space \(\mathcal H_d\otimes\mathcal H_d\), also known as Liouville space, with the inner product \((X|Y):=\mathrm{tr}\!\left(X^{\dagger}Y\right)\)~\cite{EJP2020}. The operators acting on Liouville space are called superoperators.

The state of a quantum system is represented by a the density matrix \(\rho\). The vectorized form of $\rho=\sum_{i j}\rho_{ij} |i\rangle\!\langle j|$ is $|\rho)=\sum_{i j}\rho_{ij}\,|ji).$ Note that \((\rho|\rho)=\mathrm{tr}(\rho^{2})\neq 1\) in general. Hence, we define the corresponding normalized vector as $|\tilde{\rho})=\frac{|\rho)}{\sqrt{\mathrm{tr}(\rho^{2})}}.$

In Liouville space, arbitrary CPTP dynamics is described by the following master equation\cite{arXiv2406.08584}:
\begin{equation}
\frac{d}{dt}\,|\tilde{\rho}_t)=
\left(\mathcal L_t - \frac{1}{2}\Big\{\,(\tilde{\rho}_t|\mathcal L_t|\tilde{\rho}_t)+(\tilde{\rho}_t|\mathcal L_t^{\dagger}|\tilde{\rho}_t)\,\Big\}\right)\,|\tilde{\rho}_t),
\label{f1}
\end{equation}
where $\mathcal{L}_{t}$ is the generator of the dynamics (the Liouvillian superoperator) and $\mathcal{L}_{t}^{\dagger}$ is its adjoint. The normalized state vector corresponding to the time--evolved density matrix $\rho_t$ is
\[
|\tilde{\rho}_t):=\frac{|\rho_t)}{\lVert\rho_t\rVert},\qquad
\lVert\rho_t\rVert =\sqrt{\mathrm{tr}\!\left(\rho_t^{\,2}\right)},
\]
which is the Hilbert--Schmidt norm of \(\rho_t\).
For the rest of the article, we drop the subscript \(t\) in \(\mathcal L_t\) and assume that the Liouvillian is time dependent unless stated otherwise. 

Ref.~\cite{arXiv2406.08584} provides the definition of the Liouville--space angle. In this work we mainly use the Bures angle, which is related to the geometric imaginarity measure.
The Liouville--space angle  and the  Bures angle between two states $\rho,\sigma$ are
\[
\Theta_L(\rho,\sigma):=\arccos(\tilde\rho|\tilde\sigma),\qquad
\Theta_B(\rho,\sigma):=\arccos\sqrt{F(\rho,\sigma)},
\]
where $(\tilde\rho|\tilde\sigma)=\frac{\mathrm{tr}(\rho\sigma)}{\sqrt{\mathrm{tr}(\rho^2)\mathrm{tr}(\sigma^2)}}$
and $\sqrt{F(\rho,\sigma)}=\mathrm{tr}\sqrt{\sqrt\rho\,\sigma\,\sqrt\rho}$.

To connect the two angles introduced above, we first compare the Bures angle $\Theta_B$ with the Liouville--space angle $\Theta_L$.  The next lemma shows that  $\Theta_B$ never exceeds $\Theta_L$.  This comparison will be used later to translate Liouville--space bounds into statements for geometric imaginarity.
\begin{lemma}\label{lem:Bures_vs_Liouville}
For all density operators $\rho,\sigma$,
\begin{equation}
\sqrt{F(\rho,\sigma)}\ \ge\ \frac{\mathrm{tr}(\rho\sigma)}{\sqrt{\mathrm{tr}(\rho^2)\mathrm{tr}(\sigma^2)}} \;=\;(\tilde\rho|\tilde\sigma).
\label{lemma2}
\end{equation}
Consequently, $\Theta_B(\rho,\sigma)\le \Theta_L(\rho,\sigma)$.
\end{lemma}

$\mathit{Proof}$.
By Schatten norm monotonicity, $\|X\|_1\ge \|X\|_2$ for $X=\sqrt\rho\,\sqrt\sigma$. Hence,
$\sqrt{F(\rho,\sigma)}=\|X\|_1\ge \|X\|_2=\sqrt{\mathrm{tr}(\rho\sigma)}$.
By the Cauchy--Schwarz  inequality for the Hilbert--Schmidt inner product,
$\mathrm{tr}(\rho\sigma)\le \sqrt{\mathrm{tr}(\rho^2)\mathrm{tr}(\sigma^2)}$.
We have $\sqrt{x}\ge x/\sqrt{M}$ for any $0\le x\le M$.
Taking $x=\mathrm{tr}(\rho\sigma)$ and $M=\mathrm{tr}(\rho^2)\mathrm{tr}(\sigma^2)$ we obtain
$\sqrt{\mathrm{tr}(\rho\sigma)}\ge \mathrm{tr}(\rho\sigma)/\sqrt{\mathrm{tr}(\rho^2)\mathrm{tr}(\sigma^2)}$,
which gives \eqref{lemma2}. Since $\arccos(\cdot)$ is decreasing on $[0,1]$,
$\Theta_B\le \Theta_L$ follows.


With this static comparison in hand, we now turn to dynamics.  Working in Liouville space with the normalized vectors $|\tilde\rho_t)$, we derive a time--averaged bound on how fast the state can evolve, expressed in terms of the variance (fluctuation) of the Liouvillian along the trajectory.  Combined with Lemma~\ref{lem:Bures_vs_Liouville}, this bound can be converted into a constraint on the change of geometric imaginarity.

{\textbf{The Proof of Theorem 4}.} The norm--preserving evolution of $|\tilde\rho_t)$ obeys
\[
\frac{d}{dt}|\tilde\rho_t) \;=\;
\Big(\mathcal{L}_t-\tfrac12\{(\tilde\rho_t|\mathcal{L}_t|\tilde\rho_t)+(\tilde\rho_t|\mathcal{L}_t^\dagger|\tilde\rho_t)\}\Big)\,|\tilde\rho_t).
\]
Let $P_t:=|\tilde\rho_t)(\tilde\rho_t|$ and $P_0:=|\tilde\rho_0)(\tilde\rho_0|$.
Following the steps of Theorem~1 in Ref.~\cite{arXiv2406.08584} one obtains
\begin{equation}\label{eq:rate_bound}
\left|\frac{d}{dt}\,\mathrm{tr}(P_tP_0)\right|
\;\le\; 2\,\Delta\mathcal{L}_t.
\end{equation}
Using $\left|\int_0^T f(t)\,dt\right|\le \int_0^T |f(t)|\,dt$ and integrating
\eqref{eq:rate_bound} in the standard way (divide both sides by
$\sqrt{\mathrm{tr}(P_tP_0)}\sqrt{1-\mathrm{tr}(P_tP_0)}$ and integrate) yields
\[
\Theta_L(\rho_0,\rho_T) \;\le\; \int_0^T\!dt\,\Delta\mathcal{L}_t
\;=\; T\,\Lambda_{\Delta\mathcal{L}}(T),
\]
which gives
\begin{equation}\label{eq:Liouville_QSL}
T \;\ge\; \frac{\Theta_L(\rho_0,\rho_T)}{\Lambda_{\Delta\mathcal{L}}(T)}.
\end{equation}
Finally, by Lemma~\ref{lem:Bures_vs_Liouville} and the geometric-imaginarity relations
derived in the main text  Eqs.\eqref{theta1}--\eqref{ineq2},
$|\Delta_I|\le \Theta_B(\rho_0,\rho_T)\le \Theta_L(\rho_0,\rho_T)$, which proves Theorem 4.

\end{document}